\documentclass[journal,10pt,onecolumn,draftclsnofoot,]{IEEEtran}

\usepackage{graphicx} 
\usepackage{float}
\usepackage{mathtools}
\usepackage{amsmath}
\usepackage{amsthm}
\usepackage{amssymb}
\usepackage{graphicx}
\usepackage{tikz,pgfplots}     
\usepackage{cuted}
\usepackage{verbatim}
\usepackage{xcolor}
\usepackage{mathtools}
\usepackage{dsfont}
\usepackage{bbm}
\usepackage{mathrsfs}
\usepackage{enumerate}
\pgfplotsset{compat=1.15}
\usepackage{mathrsfs}
\usetikzlibrary{arrows}
\usetikzlibrary{decorations.pathreplacing,calligraphy}
\usetikzlibrary{shapes.geometric}
\usetikzlibrary{graphs}
\usepackage{tikz-qtree}
\usepackage{filecontents}
\usepackage{csvsimple}
\usetikzlibrary{arrows.meta}
\usetikzlibrary{pgfplots.groupplots}
\usepackage[normalem]{ulem}
\usepackage{url}
\usepackage{algorithm}
\usepackage{algpseudocode}
\usepackage{microtype}

\newtheorem{definition}{Definition}
\newtheorem{lemma}{Lemma}

\newtheorem{proposition}{Proposition}

\newtheorem{remark}{Remark}
\newtheorem{theorem}{Theorem}

\newcommand{\indicator}[1]{\mathbbm{1}_{#1}}

\newcommand{\noiselessN}{{{NL}$_{[N]}$}}

\newcommand{\noiselessNi}{{{NL}$_{[N_i]}$ }}

\newcommand{\bupcnq}{{$\Pi_{n}^q$ }}
\newcommand{\bupcq}{{$\Pi^q$ }}
\newcommand{\bupcnsq}{{$\{\Pi^q_{n}\}_{n \in \bbN}$ }}

\newcommand{\numblocksi}{{l_i}}
\newcommand{\numblocksfixed}{{l}}

\def\bx{\mathbf{x}}

\def\cA{\mathcal{A}}
\def\cB{\mathcal{B}}
\def\cP{\mathcal{P}}
\def\cD{\mathcal{D}}
\def\cT{\mathcal{T}}
\def\cU{\mathcal{U}}
\def\cG{\mathcal{G}}
\def\cZ{\mathcal{Z}}

\def\cN{\mathcal{N}}

\def\allones{{\indicator{}}}

\def\errorpower{\mu}

\def\intsec{{\frac{\Delta}{\Gamma}}}

\def\bv{\mathbf{v}}
\def\vx{\mathbf{x}}
\def\vy{\mathbf{y}}
\def\vv{\mathbf{v}}
\def\vf{\mathbf{f}}

\def\mbP{\mathbb{P}}

\def\vY{\mathbf{Y}}

\def\uvy{\underline{\vy}}

\def\bbN{\mathbbm{N}}

\def\cTn{\mathcal{T}^{(n)}_q}
\def\cAq{\mathcal{A}_{q}}
\def\cAqn{\mathcal{A}_{q}^{n}}
\def\cTxn{\cT_{\bx}^{(n)} }
\def\Rni{R_{i}}

\def\bfpcq{\Pi_{n}^q}
\def\numblocksfixed{l}

\def\vecx{\underline{\vx}}
\def\vecy{\underline{\vy}}

\def\mbQ{\mathbb{Q}}

\def\ajk{a_{\{j,k\}}}
\def\a12{a_{\{1,2\}}}

\def\nlli{{\text{NL}_{[N^{l_i}]} }}
\def\nll{\text{NL}_{[N^l]}} 

\def\cTqn{\cT_{[q]}^{(n)}}
\def\cTn{\cT^{(n)}}

\newcommand\blfootnote[1]{%
  \begingroup
  \renewcommand\thefootnote{}\footnote{#1}%
  \addtocounter{footnote}{-1}%
  \endgroup
}

\begin{document}

\title{Identification over Permutation Channels}

\author{
  \IEEEauthorblockN{Abhishek~Sarkar and Bikash~Kumar~Dey}\\
 \IEEEauthorblockA{Department of Electrical Engineering\\
                    Indian Institute of Technology Bombay\\
                    \{absarkar, bikash\}@ee.iitb.ac.in}
}
 \maketitle
 \pagenumbering{arabic}

\thispagestyle{empty}

\begin{abstract}\blfootnote{Part of this work has been accepted for presentation in IEEE Information Theory Workshop, 2024.}
We study message identification over a $q$-ary uniform permutation channel, where the transmitted vector is permuted by a permutation chosen uniformly at random. For discrete memoryless channels (DMCs), the number of identifiable messages grows doubly exponentially. Identification capacity, the maximum second-order exponent, is known to be the same as the Shannon capacity of the DMC. Permutation channels support reliable communication of only polynomially many messages. A simple achievability result shows that message sizes growing as $2^{\epsilon_nn^{q-1}}$ are identifiable for any $\epsilon_n\rightarrow 0$. We prove two converse results. A ``soft'' converse shows that for any $R>0$, there is no sequence of identification codes with message size growing as $2^{Rn^{q-1}}$ with a power-law decay ($n^{-\mu}$) of the error probability. We also prove a ``strong" converse showing that for any sequence of identification codes with message size $2^{R_n n^{q-1}}$, where $R_n \rightarrow \infty$, the sum of type I and type II error probabilities approaches at least $1$ as $n\rightarrow \infty$. To prove the soft converse, we use a sequence of steps to construct a new identification code with a simpler structure which relates to a set system, and then use a lower bound on the normalized maximum pairwise intersection of a set system. 
To prove the strong converse, we use results on approximation of distributions.
The achievability and converse results are generalized to the case of coding over multiple blocks. We finally study message identification over a $q$-ary uniform permutation channel in the presence of causal block-wise feedback from the receiver, where the encoder receives an entire $n$-length received block after the transmission of the block is complete. We show that in the presence of feedback, the maximum number of identifiable messages grows doubly exponentially and we present a two-phase achievability scheme.

\end{abstract}

\section{Introduction}

In reliable communication over a channel, the message is encoded into a vector of transmit symbols, and the decoder outputs a message based on the received vector. The number of messages that can be reliably communicated  grows exponentially with the block length. The maximum possible exponential rate of transmission under an arbitrarily low probability of error is known as the Shannon capacity ($C$) of the channel.

We consider a uniform permutation channel over a $q$-ary alphabet $\cAq$, where the components of the transmitted $n$-length vector $\vx\in\cAqn$ are permuted by the channel using a permutation chosen uniformly at random from $S_n$, the set of permutations of $\{1,2,\cdots,n\}$. Since the transmit vectors of the same type are not distinguishable, it is easy to show that the maximum number ($M$) of messages that can be reliably transmitted is the number of types $N$ over the channel input alphabet; this number grows polynomially with blocklength $n$. Thus the Shannon capacity is zero.

J\`{a}J\`{a}~\cite{ja1985identification} introduced the problem of message identification, where the decoder is interested in knowing whether a particular message, which is unknown to the encoder, was transmitted. 
It needs to be ensured that both the false negative (type I)
and false positive (type II) error probabilities should be small. Message identification is a weaker requirement than decoding, and here the decision regions for different messages are allowed to overlap. It was proved by Ahlswede and Dueck~\cite{ahlswede1989identification} that by using stochastic encoding, a doubly exponential number of messages can be identified for DMCs. The identification capacity, defined as the maximum achievable second order rate, $\frac{1}{n}\log\log M$, was proved to be the same as the Shannon capacity $C$ of the DMC. 
A ``soft" converse was proved. A strong converse was later proved by Han and Verd{\'u}~\cite{han1992new,han1993approximation} (also see Steinberg~\cite{steinberg1998new} and Watanabe~\cite{watanabe2021minimax}).

Proposition 1 (a version of Gilbert's bound) in~\cite{ahlswede1989identification} guarantees the existence of set systems with bounded pairwise intersections, and this played a key role in the proof of their achievability result. 

\begin{proposition}\cite[Proposition 1]{ahlswede1989identification} \label{prop:ahlswede}
For any finite set $\cZ$, $\lambda\in (0,0.5)$, and $\epsilon>0$ such that
\begin{align}
    \lambda \log \left(\frac{1}{\epsilon}-1\right) >2 \text{ and } \epsilon <\frac{1}{6}, \label{eq:epscondition}
\end{align}
 there exist $M$ subsets $\cU_1,\cdots,\cU_M\subseteq \cZ$, each of size $\epsilon|\cZ|$, such that $|\cU_i\cap \cU_j|\leq \lambda\epsilon |\cZ|$ $\forall i\neq j$ and
\begin{align}
    M \geq |\cZ|^{-1} 2^{\epsilon |\cZ|-1}.
\end{align}
\end{proposition}

The achievability proof in~\cite{ahlswede1989identification} considers the set $\cZ$ to be the set of
about $2^{nC}$ messages, which can be communicated reliably. For a uniform permutation channel, since $\log N$ bits can be reliably communicated, one may expect that $2^{\epsilon N}$ messages may be identified. However, condition \eqref{eq:epscondition} implies that for the probability of error $\lambda_n$ of a sequence of codes to go to $0$, $\epsilon_n$ must also go to $0$. Our soft converse in this paper shows that $2^{\epsilon N}$ (where $\epsilon$ is fixed) messages cannot be identified via uniform permutation channel with probability of error decaying as a power of $n$.

We prove an achievability and two converse results for identification over uniform permutation channels over the alphabet $\cAq$. Proposition~\ref{prop:ahlswede} is used to get the achievability result showing that any message size growing as $2^{\epsilon_n n^{q-1}}$, where $\epsilon_n\rightarrow 0$, is identifiable. We prove a soft converse showing that for any $R,\mu>0$, no sequence of ID codes exist with message size growing as $2^{Rn^{q-1}}$ and probability of errors vanishing as $n^{-\mu}$. The soft converse meets the achievability result in terms of the message size, but it rules out only a power-law decay of the probability of errors. We also prove a strong converse showing that for any $R_n \rightarrow \infty$, and a sequence of ID codes with message size growing as $2^{R_n n^{q-1}}$, the sum of type I and type II probability of errors approaches at least $1$. While the strong converse gives a stronger guarantee on the probability of error, it gives a weaker bound on the message size.

We also consider coding over multiple ($l$)  uses of the permutation channel, which amounts to the transmission of $ln$ symbols. The channel uniformly permutes each block of $n$ symbols independently. For the reliable communication problem, since one out of $N$ messages can be communicated in one block with zero probability of error, coding over multiple blocks does not give any advantage either in terms of message size or probability of error. Over $l$ blocks, it is possible to transmit one out of $N^l$ messages with zero probability of error. Hence it is expected that the identifiable message size is roughly of the order of $2^{N^l}$. However, the number of identifiable messages depends on how $l$ increases in comparison with $n$. Please see Theorem~\ref{thm:qary_multishot} and the discussion preceding the theorem.

We next consider block-wise perfect causal feedback to the encoder. For a permutation channel, symbol-wise causal feedback leads to a paradoxical situation where the encoder knows a transmitted symbol before encoding the very same symbol, resulting in an ill-defined operation. In block-wise causal feedback, the encoder gets the full $n$-length received vector only after the corresponding block has been transmitted. We consider feedback encoding over multiple ($\numblocksfixed$) blocks and study the growth rate of the number of identifiable messages as $n$ increases. We show that the number of messages grows doubly exponentially in $nl$ (as $2^{q^{Rnl}}$ with $\sup R = 1$), as long as $n \rightarrow \infty$.
Our achievability scheme uses a two-phase protocol, similar to \cite{ahlswede1989identification_f}. In the first phase, the transmitter and receiver establish some common randomness. This common randomness, along with the message, is then utilized to determine the $n$-length vector to be transmitted in the next phase.

After the identification problem was introduced in~\cite{ja1985identification,ahlswede1989identification} it has been studied in various setups. We now give a glimpse of such works, without being exhaustive. Identification over DMCs with feedback under both deterministic and stochastic encoding was studied in~\cite{ahlswede1989identification_f,ahlswede1995new}, while identification over wiretap channels was studied in~\cite{spandri2023information}, \cite{rosenberger2023capacity},~\cite{ahlswede1995new}.
Identification over broadcast channels was investigated in ~\cite{bracher2017identification},~\cite{rosenberger2023identification}, while identification in the presence of feedback over multiple-access channels and broadcast channels was studied in~\cite{ahlswede1991identification}.
Identification was studied over Gaussian channels in~\cite{salariseddigh2021deterministic},~\cite{labidi2021identification}; over additive noise channels under average and peak power constraints in~\cite{wiese2022identification}; over compound channels and arbitrarily varying channels in~\cite{ahlswede1999identification}. 
Deterministic identification was studied for DMCs with and without input constraints in~\cite{salariseddigh2021deterministic_it}, and for discrete-time Poisson channels with constraints on arrival rates in~\cite{salariseddigh2023deterministic}. 

The permutation channel is relevant in DNA-based storage systems,  multipath routing of packets in communication networks, and diffusion-based molecular communication~\cite{makur2020coding}.
In multipath routed networks, in the absence of any packet id in the packets, the out-of-order arrival of packets due to varying delay or changing topologies may be thought to be due to random permutations~\cite{gadouleau2010binary, walsh2009optimal}.
The studies in~\cite{kovavcevic2015perfect, kovavcevic2018codes} examined coding in channels with random permutations and other impairments such as insertion, deletion and substitution.  See~\cite{akan2016fundamentals} and~\cite{pierobon2014fundamentals} for a comprehensive survey of molecular communication systems and the role of the permutation channel in diffusion-based communication systems.
An overview of coding challenges for DNA-based storage is presented in~\cite{Sima2023ErrorCF}, while~\cite{sima2023robust} presents an optimal code construction for correcting multiple errors in unordered string-based data encoding within DNA storage systems. See~\cite{shomorony2022information} for a comprehensive study of DNA-based storage systems. 
\cite{langberg2017coding,chee2016string} examined coding for permutation channels with restricted movements. Noisy permutation channels have been studied in~\cite{makur2020coding,tang2023capacity, tandon2019bee,tandon2020bee}. To the best of our knowledge, 
identification over permutation channels has not been studied before.

The organization of this paper is as follows.
We present the problem setup in Sec.~\ref{sec:system_model}, and the main results in Sec.~\ref{sec:results}. In Sec.~\ref{sec:mod_ID_Codes_q}, we give a series of constructions through which we derive a new `canonical' code from a given identification code. In Sec.~\ref{sec:set_sysq}, we prove a bound on the intersection of a set system. The results from Sections~\ref{sec:mod_ID_Codes_q} and \ref{sec:set_sysq} are later used in the proofs of the soft converses. The proof of  Theorem~\ref{thm:oneshot} (on one-shot identification) is presented in Sec.~\ref{sec:pf_thm_2A}. The proof of  Theorem~\ref{thm:qary_multishot} (on multi-shot identification) is presented in Sec~\ref{sec:multishot}.  The proof of  Theorem~\ref{thm:feedback_deterministic} (on identification with feedback) is presented in Sec~\ref{sec:feedback}. 
We conclude the paper in Sec.~\ref{sec:conclusion}.

\section{Problem Statement}\label{sec:system_model}

For a positive integer $k$, we denote $[k]:=\{1,\cdots,k\}$. For any set $\cA$, $\cP(\cA)$ denotes the set of all probability mass functions over  $\cA$. Let $\cAq$ be a set of $q$ elements
and $\cTqn \subset \cP(\cAqn)$ be the set of all types over $\cAqn$ and denote $N \coloneqq \left|\cTqn\right|=\binom{n+q-1}{q-1}$. We note that
\begin{align} 
    \frac{n^{(q-1)}}{(q-1)!} \le N &\le \frac{n^{(q-1)}}{(q-1)!}\left( 1+\frac{q-1}{n} \right)^{q-1}\label{eq:Nbounds}&\\ 
    &\le(2n)^{(q-1)} &\text{ (for $n\ge q-1$)}. \label{eq:Nbounds1}
\end{align}
For convenience, we index the types in $\cTqn$ by $[N]$ and denote $\cTqn=\{T_1,T_2,\cdots,T_N\}$. For a type $P \in \cTqn$, the type class of $P$, denoted by $\cT_{P}^{(n)}$, is the set of vectors in $\cAq^n$ with type $P$.
For two sets $\cA,\cB$, we denote by $\cP(\cA|\cB)$ the set of all conditional distributions $P_{X|Y}$ where $X\in\cA, Y\in\cB$. 
Any channel with input alphabet $\cA$ and output alphabet $\cB$ is specified by a conditional probability distribution $P\in \cP(\cB|\cA)$. The ``noiseless channel'' with input and output alphabets $\cA$, where the output is the same as the input, is denoted by NL$_{\cA}$. In particular, \noiselessN~denotes the noiseless channel with input and output alphabets $[N]$. We now consider $q$-ary permutation channels. Without loss of generality, consider the $q$-ary alphabet $\cAq:=[q]$. For any $\vx\in \cAqn$ and $\sigma\in S_n$, we define
\begin{align*}
\sigma\vx := (x_{\sigma^{-1}(1)}, x_{\sigma^{-1}(2)}, \cdots, x_{\sigma^{-1}(n)}).
\end{align*}
The $n$-block  $q$-ary uniform permutation channel  \bupcnq is defined as the channel with input and output alphabets $\cAqn$ and transition probability
\begin{align} \label{eq:bupc transition prob}
    \bfpcq(\vy|\vx) = \frac{1}{n!}\sum_{\sigma \in S_n} \indicator{\left\{\vy=\sigma\vx \right\}}
\end{align}
for all $\vx,\vy \in \cAqn$.
We will refer to the sequence of channels \bupcnsq as the uniform permutation channel over $\cAq$ and denote it by \bupcq\!\!. 
For every $\vx \in \cAqn$, the action of permutations generate the typeclass of $\vx$:
\begin{align*}
\cTxn \coloneqq \{\sigma \vx| \sigma \in S_n\},
\end{align*} 
In the language of group action, $\cTxn$ is the orbit of $\vx$ under the action of $S_n$ on $\cAqn$.  It is easy to check that if $\vx$ is transmitted and a permutation $\sigma\in S_n$ is chosen uniformly at random, then the output $\sigma(\vx)$ is uniformly distributed in $\cTxn$. 

\subsection{Identification codes (one-shot)}
We now define identification codes over a single use of an arbitrary channel. The encoders are defined to be stochastic, as it is known that in absence of feedback, the number of messages that can be identified using deterministic encoders is the same as the number of messages that can be communicated reliably~\cite{koga2002information}.
\begin{definition}\label{def:code}
	An $M$-ary identification (ID) code with deterministic decoders  for any channel $P\in \cP(\cB|\cA)$ is a set
	\begin{align*}
		\left\{(Q_i,\cD_i)\mid i=1,\ldots,M\right\}
	\end{align*}
	of pairs with
	$
		Q_i \in \cP(\cA),
	$  and $\cD_i \subset \cB,$
for $i=1,\ldots,M$. For such a code, a message $i$ is encoded to a symbol
$x\in\cA$ with probability $Q_i(x)$, and the decoder for message $i$ outputs $1$ (``Accept'') if the received symbol
$y\in \cD_i$, and outputs $0$ (``Reject'') otherwise. If, for each $i$, the encoding distribution $Q_i$ is the uniform distribution over some support set $\cA_i$, then the ID code  is specified by $\left\{(\cA_i,\cD_i)\mid i=1,\ldots,M\right\}$.
\end{definition}

\begin{definition}\label{def:stochastic ID code}
	An $M$-ary identification (ID) code with stochastic decoders  for any channel $P\in \cP(\cB|\cA)$ is a set
	\begin{align*}
		\left\{(Q_i,P_i)\mid i=1,\ldots,M\right\}
	\end{align*}
	of pairs with
	$
		Q_i \in \cP(\cA),
	$  and $P_i \in \cP(\{0,1\}|\cB),$
for $i=1,\ldots,M$. For such a code, when $y$ is received, the decision rule $P_i$ for message $i$ outputs $1$ with probability $P_i(1|y)$ and $0$ with probability $P_i(0|y)$. Clearly, an ID code with deterministic decoders is a special case with $P_i(1|y):=\indicator{\{y\in \cD_i\}}$.
\end{definition}

While it is known that stochastic decoders do not help in increasing the number of identifiable messages~\cite{ahlswede1989identification, ahlswede1999identification} (also see Lemma~\ref{lemma: permutation to noiseless_multishot} and the discussion preceding it), stochastic decoders will appear in an intermediate step in the proofs of our soft converses. For an ID code with stochastic decoders, the probability that a message $j\neq i$ is accepted (``false alarm'' for $j$) when $i$ is encoded is given by
\begin{align} 
    \lambda_{i \rightarrow j} &\coloneqq 
    \sum_{x \in \cA, y\in \cB}Q_i(x)P(y|x)P_j(1|y) \label{eq:error id at n}& \\
    & = \sum_{x \in \cA, y\in \cD_j}Q_i(x)P(y|x) \,&\text{ (for deterministic decoders)} 
\end{align}
Similarly, the missed detection probability for message $i$, i.e. the probability that the decoder rejects message $i$ while $i$ was encoded, is given by
\begin{align}
    \lambda_{i \not\rightarrow i}& \coloneqq 
    \sum_{x \in \cA, y\in \cB}Q_i(x)P(y|x)P_i(0|y) \label{eq:error id at n_2}& \\
    & = \sum_{x \in \cA, y\not\in \cD_i}Q_i(x)P(y|x) \,&\text{ (for deterministic decoders)}
\end{align}
We define the Type-I and Type-II probability of errors of the code as
\begin{align} 
\lambda_1 \coloneqq \max_{1 \le i\le M} \lambda_{i \not\rightarrow i}, \hspace{10mm}
&\lambda_2 \coloneqq \max_{1\le i\neq j\le M}\lambda_{i \rightarrow j},
\end{align}
And we define the sum probability of error as
\begin{align}  \label{eq:ID error}
    \lambda :=\lambda_1+\lambda_2.
\end{align}
Now consider the permutation channel \bupcnq\!. Note that a single use of this channel involves the transmission of $n$ symbols from $\cAq$. Hence a one-shot code over \bupcnq transmits $n$ symbols from $\cAq$.
We refer to an $M$-ary (i.e. with $M$ messages) ID code for \bupcnq as an $(n,M)$ ID code (or $(n,M,\lambda_1,\lambda_2)$ ID code) for the uniform permutation channel \bupcq (or \bupcnq\!), and similarly, an ID code for \noiselessN~is referred as an $(N,M)$ ID code (or $(N,M,\lambda_1,\lambda_2)$ ID code) for \noiselessN. 
Unless otherwise specified, an ID code will refer to one with deterministic decoders throughout the paper. For an $(n,M)$ ID code with deterministic decoders for \bupcq\!\!, the probability of errors are given by
\begin{align}
    \lambda_{i \rightarrow j} &\coloneqq 
    \sum_{\bx \in \cAqn}Q_i(\bx)\frac{|\cTxn \cap \cD_j|}{|\cTxn|}. \label{eq:error id at n_2_1}
\end{align}
and
\begin{align}
    \lambda_{i \not\rightarrow i}\coloneqq 
    \sum_{\bx \in \cAqn}Q_i(\bx)\frac{|\cTxn \cap \mathcal{D}_i^c|}{|\cTxn|}\label{eq:error id at n_1}.
\end{align}

\subsection{Multi-shot identification codes}
We now consider a communication scheme over $\numblocksfixed$ uses of the channel \bupcnq\!,  
i.e. over the transmission of $\numblocksfixed$ blocks of $n$ symbols each. The permutation of the transmitted symbols is restricted within each block. The channel \bupcnq also acts on each block memorylessly. That is, the permutations chosen over different blocks are independent.
\begin{definition}[$l$-shot identification code]
An  $\numblocksfixed$-shot $M$-ary identification code, or an $(\numblocksfixed, M)$ identification code, with deterministic decoders over a channel 
$P\in \cP(\cB|\cA)$ is a set
	\begin{align*}
		\left\{(Q_i,\cD_i)\mid i=1,\ldots,M\right\}
	\end{align*}
	of pairs with
	$
		Q_i \in \cP(\cA^{\numblocksfixed}),
	$  and $\cD_i \subset \cB^{\numblocksfixed},$
for $i=1,\ldots,M$. Similarly, an $(\numblocksfixed, M)$ identification code with stochastic decoders over a channel 
$P\in \cP(\cB|\cA)$ is defined as a set 
	\begin{align*}
		\left\{(Q_i,P_i)\mid i=1,\ldots,M\right\}
	\end{align*}
	of pairs with
	$
		Q_i \in \cP(\cA^{\numblocksfixed}),
	$  and $P_i \in \cP(\{0,1\}|\cB^{\numblocksfixed}),$
for $i=1,\ldots,M$. Unless otherwise specified, a multi-shot ID code will refer to one with deterministic decoders.
\end{definition}
The probabilities of different types of errors for multi-shot codes are defined in a similar manner as in the case of one-shot codes, by replacing $\cA$ with $\cA^l$ and $\cB$ with $\cB^l$ in \eqref{eq:error id at n}-\eqref{eq:ID error}. We refer to an $l$-shot $M$-ary (i.e. with $M$ messages) ID code for \bupcnq as an $(n, l, M)$ ID code (or $(n,l, M,\lambda_1,\lambda_2)$ ID code) for the uniform permutation channel \bupcq (or \bupcnq\!).

\subsection{Permutation channel with noiseless feedback} 

We first define an identification-feedback code for any channel $P \in \cP(\cB|\cA)$ with symbol-wise feedback.
Let  $\vecx=(x^{(1)},\ldots, x^{(\numblocksfixed)})$  denote the transmitted sequence of $\numblocksfixed$ symbols. Similarly, let $\vecy=(y^{(1)},\ldots, y^{(\numblocksfixed)})$ denote the received sequence of $\numblocksfixed$ symbols. 

\begin{definition}
    An $(\numblocksfixed,M)$ identification-feedback (IDF) code with stochastic encoder for a channel $P \in \cP(\cB|\cA)$ is a set 
    \begin{align}
        \{(\mbQ_{i,1}, \cdots, \mbQ_{i,\numblocksfixed},\cD_i)|i=1,\ldots,M\}
    \end{align}
    where, for every message $i\in [M]$ and every block $j=[\numblocksfixed]$,  
    $\mbQ_{i,j} \in  \cP\left(\cA|\cB^{j-1} \times \cA^{j-1}  \right)
    $ are the encoding conditional distributions, and  
     $\cD_i \subset \cB^{\numblocksfixed}$ are the decoding sets for all $i \in [M]$.
     For message $i$ and  given the feedback 
     $(y^{(1)},\ldots,y^{(j-1)}) \in \cB^{j-1} $ and all past encoded symbols $( x^{(1)}, \ldots, x^{(j-1)})$, 
     in block $j$ the encoder generates $x^{(j)}$ using the distribution $\mbQ_{i,j}(x^{(j)}| y^{(1)},\ldots,y^{(j-1)},  x^{(1)}, \ldots, x^{(j-1)})$. 
     The decoder for message $i$ outputs $1$ (``Accept") if the received vector $(y^{(1)},\ldots, y^{(\numblocksfixed)}) \in \cD_i$ and outputs $0$ (``Reject") otherwise. 
     An IDF code with deterministic encoding functions $\vf_{i,j}$s  is a special case of stochastic  IDF codes, where these conditional distributions are of the form
     $\mbQ_{i,j}(x^{(j)}|y^{(1)},\ldots,y^{(j-1)}, x^{(1)}, \ldots, x^{(j-1)})= \indicator{\{ x^{(j)}=\vf_{i,j}(y^{(1)},\ldots,y^{(j-1)}, x^{(1)}, \ldots, x^{(j-1)})\}},
     $ where for $i\in [M], j \in [l]$, $\vf_{i,j}$s  are encoder mappings
     $\vf_{i,j}:  \cB^{j-1} \times \cA^{j-1} \rightarrow \cA$ (for $j=1$, 
     $\vf_{i,1}$ is an element of $\cA$). 
     As in Definition~\ref{def:stochastic ID code}, an IDF code with stochastic decoders is defined by replacing the decoding sets $\cD_i$ with conditional distributions $P_i\in \cP(\{0,1\}|\cB^{\numblocksfixed})$. 
\end{definition}

The transmitted and received blocks in the $j$-th use of the channel $P \in \cP(\cB|\cA)$ are denoted by $x^{(j)}$ and $y^{(j)}$ respectively and we denote $\vecx=(x^{(1)}, \ldots, x^{(\numblocksfixed)})$ and $\vecy=(y^{(1)}, \ldots, y^{(\numblocksfixed)})$.  
When message $i$ is encoded, the transmitted and received blocks have the joint distribution
\begin{align}
    &\hspace*{-5mm}\mbP^{(i)}_{\vecx,\vecy}(x^{(1)},\ldots, x^{(\numblocksfixed)}, y^{(1)},\ldots, y^{(\numblocksfixed)})\notag\\
    &= 
    \prod_{j=1}^{j=\numblocksfixed}\mbQ_{i,j}(x^{(j)}|y^{(1)},\ldots,y^{(j-1)},  x^{(1)}, \ldots, x^{(j-1)}) \notag 
    \cdot
    \mbP(y^{(j)}|y^{(1)},\ldots,y^{(j-1)},  x^{(1)}, \ldots, x^{(j)})\\
    &\stackrel{}{=}\left(\prod_{j=1}^{j=\numblocksfixed}\mbQ_{i,j}(x^{(j)}| y^{(1)},\ldots,y^{(j-1)},  x^{(1)}, \ldots, x^{(j-1)}) \right)\cdot \left(\prod_{j=1}^{j=\numblocksfixed} P (y^{(j)}|x^{(j)}) \right). \label{eq:feedback joint dist1}
\end{align}
where the last step follows from the assumption of memorylessness of the channel across blocks.
Using \eqref{eq:feedback joint dist1}, the probability that a message $j \neq i$ is accepted when $i$ is encoded is given by 
\begin{align}
\lambda_{i \rightarrow j} &= \sum_{(y^{(1)}, \ldots, y^{(\numblocksfixed)}) \in \cB^{l}} \mbP^{(i)}_{\vecy}(y^{(1)}, \ldots, y^{(\numblocksfixed)}) P_j(1|(y^{(1)}, \ldots, y^{(\numblocksfixed)})) &\text{ (with stochastic decoders)},\\
    &= \sum_{(\vy^{(1)},\ldots, \vy^{(\numblocksfixed)}) \in \cD_j}\mbP^{(i)}_{\vecy}( y^{(1)},\ldots, y^{(\numblocksfixed)}) &\text{ (with deterministic decoders)}.
\end{align}
where $\mbP^{(i)}_{\vecy}$ is the marginal of $\mbP^{(i)}_{\vecx,\vecy}$ on $\vecy$.
The probability that the decoder rejects message $i$ when $i$ was transmitted, is given by
\begin{align}
    \lambda_{i \not\rightarrow i} &= \sum_{(y^{(1)}, \ldots, y^{(\numblocksfixed)}) \in \cB^{l}} \mbP^{(i)}_{\vecy}(y^{(1)}, \ldots, y^{(\numblocksfixed)}) P_i(0|(y^{(1)}, \ldots, y^{(\numblocksfixed)}))&\text{ (with stochastic decoders)}\\
    &= \sum_{(y^{(1)},\ldots, y^{(\numblocksfixed)}) \in \cD_i^c}\mbP^{(i)}_{\vecy}( y^{(1)},\ldots, y^{(\numblocksfixed)}) &\text{ (with deterministic decoders)}.
\end{align}
The Type-I, Type-II, and sum error probabilities, respectively, $\lambda_1,\lambda_2$ and $\lambda$ are defined as before. 
Lemma~\ref{lemma: st to det gen} shows that given an IDF code with stochastic encoder and decoders for a channel $P \in \cP(\cB|\cA)$; one can construct deterministic decoders for the same encoder. The sum error probability of the IDF code with deterministic decoders increase to at most $2 \sqrt{\lambda}$, where $\lambda$ is the sum-error probability of the original code. Hence, the asymptotic order of the number of messages under stochastic decoding is the same as under deterministic decoding. Therefore, without loss of generality, we consider IDF codes with deterministic decoders.

Now consider the use of \bupcnq with block-wise feedback and coding over $l$ blocks. Over different blocks, the channel acts memorylessly, and hence we can use the definitions and derivations above in this context where $\cA,\cB,P$ are replaced with $\cAq^n,~ \cAq^n$, \bupcnq respectively. $\vx^{(i)}$ and $\vy^{(i)}$ denote respectively the $i$-th transmitted and received blocks.

We refer to an $(\numblocksfixed,M)$ IDF code for \bupcnq with Type I and Type II probabilities of errors $\lambda_1$ and $\lambda_2$ respectively as an $(n,\numblocksfixed,M)$ IDF code (or  $(n,\numblocksfixed,M,\lambda_1,\lambda_2)$ IDF code) for \bupcq (or \bupcnq\!).

\section{Main results}\label{sec:results}
We first present our results for one-shot ID codes over \bupcq\!\!.

\begin{theorem}[One-shot identification over \bupcq\!\!] \label{thm:oneshot} ~~
   \begin{enumerate}[(i)]
    \item {Achievability:} For any $\epsilon_n \rightarrow 0$, there exists a sequence of $(n,2^{\epsilon_n n^{q-1}},0,\lambda_{2,n})$ ID codes for \bupcq with $\lambda_{2,n} \rightarrow 0$ as $n \rightarrow \infty$. \label{thm:ach}
       \item {\it Soft Converse: } For any given $\errorpower, R>0$, there does not exist a sequence of $(n_i,2^{Rn_i^{q-1}},\lambda_{1,i}, \lambda_{2,i})$ ID codes for the $q$-ary uniform permutation channel \bupcq  such that  $n_i\rightarrow \infty$ and probability of errors $\lambda_{1,i}, \lambda_{2,i} < {n_i}^{-\errorpower}$ for all $i$. \label{thm:soft_converse}
       \item {\it Strong Converse: } 
       For any given sequence $\{\Rni \}_{i \ge 1}$ such that $\Rni \rightarrow \infty$  and any sequence of  $(n_i,2^{\Rni n_i^{q-1}}, \lambda_{1,i}, \lambda_{2,i})$ ID codes with $n_i \rightarrow \infty$ for the $q$-ary uniform permutation channel \bupcq\!\!, $\liminf_{i\rightarrow \infty} (\lambda_{1,i}+ \lambda_{2,i}) \geq 1$.\label{thm:strong_converse}
   \end{enumerate}
\end{theorem}

We give an outline of the proof here, the formal proof can be found in Sec.~\ref{sec:pf_thm_2A}.
The proof of achievability (Theorem~\ref{thm:oneshot} (\ref{thm:ach})) is a straightforward application of Proposition~\ref{prop:ahlswede}. 

For proving the soft converse (Theorem~\ref{thm:oneshot}(\ref{thm:soft_converse})), we use a sequence of steps to derive a sequence of ID codes with a simple structure and then prove the zero-rate converse for such a sequence of codes. However, one of these steps is to reduce the encoder distributions to uniform distributions, in a similar manner as in~\cite{ahlswede1989identification}. This step increases the probability of error by a (arbitrarily small) power of $n$. As a result, this argument works if the original sequence of codes has a probability of error decaying at least as fast as $n^{-\errorpower}$ for some $\errorpower>0$. Several other steps in the argument are new, and arise due to the current channel model. 

To prove the strong converse (Theorem~\ref{thm:oneshot}(\ref{thm:strong_converse})), for a given sequence of ID codes, we first construct a sequence of ID codes with stochastic decoders for noiseless channels over types. We then use a result on the approximation of distributions and a relation of the variational distance between encoder distributions with the probability of error. The technique is similar to that in~\cite{han1992new,han1993approximation}.
Note that a direct application of the strong converse results of~\cite{han1992new,han1993approximation} only implies a zero second-order rate (since the Shannon capacity is zero), while our strong converse rules out a message size growing as $2^{R_n n^{q-1}}$ with any $R_n\rightarrow \infty$. 

\begin{remark}
    It is reasonable to expect that if $\epsilon_n$ diminishes slowly in the achievability, resulting in a faster growth of the message size, then the probability of error $\lambda_{2,n}$ will decay slowly as well. By using the bounds \eqref{eq:Nbounds},
    one can check that for the construction in the achievability proof, $\lambda_{2,n}$  is of the order
\begin{align}
\lambda_{2,n} \simeq \begin{cases} \frac{1}{\log\left(\frac{1}{\epsilon_n}\right)} & \text{ if } \frac{\epsilon_n n^{q-1}}{\log n} \rightarrow \infty, \\
\frac{1}{\log\left(\frac{n^{q-1}}{\log n}\right)} & \text{ otherwise.} 
\end{cases}
\end{align}
\end{remark}

\begin{remark}
Note that the probabilities of type-I and type-II errors for decoder $j$ (i.e. when the decoder is interested in identifying message $j$), are $\lambda_{i,j\not\rightarrow j}$ and $\max_{k\neq j}\lambda_{i,k\not\rightarrow j}$ respectively. The strong converse also holds for each decoder, i.e., in Theorem~\ref{thm:oneshot}(\ref{thm:strong_converse}), $\liminf_{i\rightarrow \infty} \max_j(\lambda_{i, j\not\rightarrow j}+ \max_{k\neq j} \lambda_{i, k\rightarrow j}) \geq 1$. This follows by replacing $(\lambda_{1,i} + \lambda_{2,i})$ with $\max_j(\lambda_{i, j\not\rightarrow j}+ \max_{k\neq j} \lambda_{i, k\rightarrow j})$ in \eqref{eq:rmk1_1}, \eqref{eq: lambda_n and TV distance}, \eqref{eq:rmk1_2}, and \eqref{eq:rmk1_3} in the proof (Sec.~\ref{sec:strong converse oneshot proof}).
\end{remark}

We now consider $l$-shot identification over \bupcnq 
and give our relevant results.
Intuitively, for the achievability, one may use a single-shot code for communication repeatedly $l_i$ times to transmit a message from a set of $N^{l_i}\approx n_i^{l_i(q-1)}$ messages. Hence the number of identifiable messages is expected to be of the order $\approx 2^{n_i^{l_i(q-1)}}$. The results in Theorem~\ref{thm:qary_multishot} agree with this intuition. However, the converse results in Theorem~\ref{thm:qary_multishot}(\ref{thm:soft_converse_multishot_i}),(\ref{thm:soft_converse_multishot_ii}),(\ref{thm:strong_converse_multishot}) give graded guarantees on the probability of error depending on how fast $\numblocksi$ increases in comparison with $n_i$. In addition, the strong converse also gives a weaker bound on the number of identifiable messages, as in the case of single shot identification (Theorem~\ref{thm:oneshot}(\ref{thm:strong_converse})).

\begin{theorem}[Multishot identification over \bupcq]\label{thm:qary_multishot}~~
\begin{enumerate}[(i)]
    \item {\it Achievability:} For any sequences $n_i\rightarrow \infty,  \epsilon_i \rightarrow 0$, and any sequence $l_i$, there exists a sequence of $(n_i,l_i,2^{\epsilon_i n_i^{l_i(q-1)}},0,\lambda_{2,i})$ ID codes for \bupcq for some $\lambda_{2,i} \rightarrow 0$. \label{thm:achv_multishot}
    \item {\it Soft Converse I: } For any given $\errorpower, R>0$, there does not exist a sequence of $(n_i,l_i,2^{Rn_i^{l_i(q-1)}},\lambda_{1,i}, \lambda_{2,i})$ ID codes for the $q$-ary uniform permutation channel \bupcq  such that  $n_i\rightarrow \infty$,  $\lambda_{1,i}, \lambda_{2,i} < {n_i}^{-\errorpower}$, and  $ l_i = o\left(n_i^{\omega}\right)$ (i.e. $\frac{l_i}{n_i^{\omega}} \rightarrow 0$) where $\omega = \min{\{1,\errorpower/4\}}$.\label{thm:soft_converse_multishot_i}
    \item {\it Soft Converse II: } For any given $\errorpower, R>0$, there does not exist a sequence of $(n_i,l_i,2^{Rn_i^{l_i(q-1)}},\lambda_{1,i}, \lambda_{2,i})$ ID codes for the $q$-ary uniform permutation channel \bupcq  such that  $n_i\rightarrow \infty$,  $\lambda_{1,i}, \lambda_{2,i} < {n_i}^{-l_i\errorpower}$, and  $l_i = o(n_i)$ (i.e. $\frac{l_i}{n_i}\rightarrow 0$ as $i\rightarrow \infty$).\label{thm:soft_converse_multishot_ii}
    \item {\it Strong Converse: } 
    For any sequences  of $(n_i,l_i,2^{\Rni n_i^{l_i(q-1)}}, \lambda_{1,i}, \lambda_{2,i})$ ID codes for the $q$-ary uniform permutation channel \bupcq with $n_i \rightarrow \infty$,  $\Rni \rightarrow \infty$,  and $l_i = O(n_i)$ (i.e. $l_i \leq \beta n_i$  for all $i$ for some constant $\beta$), $\liminf_{i\rightarrow \infty} (\lambda_{1,i}+ \lambda_{2,i}) \geq 1$.\label{thm:strong_converse_multishot}
   \end{enumerate}
\end{theorem}

Note that for $l_i=1$, each part of Theorem~\ref{thm:qary_multishot} gives the corresponding one-shot result in Theorem~\ref{thm:oneshot}. In particular, both versions of soft converse in Theorem~\ref{thm:qary_multishot}(\ref{thm:soft_converse_multishot_i}) and (\ref{thm:soft_converse_multishot_ii}) give the same soft converse for the one-shot case.

The proof of Theorem~\ref{thm:qary_multishot}
follows similar arguments as that of Theorem~\ref{thm:oneshot}, and can be found in Sec.~\ref{sec:multishot}. 

We now present our main result on identification via $q$-ary uniform permutation channel with causal block-wise feedback.
First, consider a fixed $n$ and the corresponding channel \bupcnq\!, and coding over $\numblocksfixed\rightarrow \infty$ blocks over this fixed channel \bupcnq\!. Since the channel \bupcnq is a DMC, it is known ~\cite{ahlswede1989identification_f} that the number of identifiable messages grows doubly exponentially with $\numblocksfixed$, and the largest second-order exponent of the message size is given by
\begin{align}
C_{n,FB,d}&= \max_{\vx \in \cAq^n} H(\vY|\vx) \notag\\
& = \max_{\vx \in \cAq^n} \log |\cT_{\vx}^{(n)}| \notag \\
& = n\log q - o(n).\notag 
\end{align}
under deterministic coding. Under stochastic coding (with private randomness at the encoder), the largest second-order exponent of the message size is given by
\begin{align}
C_{n,FB,s}&= \max_{P_{\vx}} H(\vY) \notag \\
& = n\log q. \notag 
\end{align}

 Hence the number of identifiable messages grows as $\approx 2^{q^{nl}}$ under both deterministic and stochastic coding. This can be directly used to conclude that for any $R<1$ and sequence $n_i\rightarrow \infty$, a sequence $l_i\rightarrow \infty$ may be chosen (so that $l_i$ is large enough) so that there is a sequence of $(n_i,l_i,2^{q^{Rn_il_i}})$ codes with vanishing probabilities of errors. The achievability in the following theorem gives a stronger guarantee that there are such codes with a {\it fixed}\footnote{It is easy to check in the proof that the achievability also holds for any sequence of $l_i>\frac{2}{1-R}$.} value of $l$.

\begin{theorem} [Identification over \bupcq with noiseless Feedback] \label{thm:feedback_deterministic}
\begin{enumerate}[(i)]~~
    \item {\it Achievability:} For any $R<1$ and $l>\frac{2}{1-R}$, there exists a sequence of $(n_i,l,2^{q^{Rln_i}},0,\lambda_{2,i})$ deterministic IDF codes for \bupcq with $n_i\rightarrow \infty$ and $\lambda_{2,i} \rightarrow 0$. \label{thm:ach_f}
    \item Converse: For any $n,l\geq 1$, and for any $(n,l,2^{q^{nl}}, \lambda_{1}, \lambda_{2})$ IDF code with stochastic (or deterministic) encoder for the $q$-ary uniform permutation channel \bupcnq\!,  $\lambda_{1}+ \lambda_{2} \geq 1$.\label{thm:converse_f}
    \end{enumerate}
\end{theorem}

The proof of Theorem~\ref{thm:feedback_deterministic} is given in Sec.~\ref{sec:feedback}. The achievability proof uses a two-phase scheme similar to that in \cite{ahlswede1989identification_f}. The first phase over $l-1$ blocks establishes a common randomness between the transmitter and receiver. The transmission in the second phase, i.e. the last block, depends on this common randomness and the message. 

It is reasonable to define the rate of an $(n,l,M)$ IDF code for \bupcnq as 
\begin{align}
    R \coloneqq \frac{1}{nl}\log_q \log M,
\end{align}
and the capacity to be the supremum of rates {\it achievable} via sequences of codes with vanishing probability of error. Theorem~\ref{thm:feedback_deterministic} shows that the IDF capacity of  \bupcq is $1$, and this can be achieved by coding over any sequence of $l_i\rightarrow \infty$ blocks.

\section{Modification Of ID Codes}\label{sec:mod_ID_Codes_q}

Towards proving the soft converses (Theorem~\ref{thm:oneshot}(\ref{thm:soft_converse}), and Theorem~\ref{thm:qary_multishot}(\ref{thm:soft_converse_multishot_i}),(\ref{thm:soft_converse_multishot_ii})), in this section, we provide five simple code modification steps through which, given an ID code for \bupcnq, we construct a code for \noiselessN~with uniform encoder distributions and equal-sized supports.
The first step, i.e., constructing an ID code for \noiselessN~with stochastic decoders from an ID code for \bupcnq\!, is also used in the proof of the strong converses (Theorem~\ref{thm:oneshot}(\ref{thm:strong_converse}) and Theorem~\ref{thm:qary_multishot}(\ref{thm:strong_converse_multishot})).

\subsection{From ID code with deterministic decoders for \bupcnq to a ID code with stochastic decoders for \noiselessN}
For every $(n, M)$ ID code with deterministic decoders for \bupcnq,
we present a construction of an $(N, M)$ ID code with stochastic decoders for \noiselessN~ with the same probability of error.

\begin{lemma}\label{lemma: permutation to noiseless}
    Given an ID code with deterministic decoders $\{(Q_i, \mathcal{D}_i) | i=1,\ldots, M\}$ for \bupcnq,
     there exists an ID code with stochastic decoders for the noiseless channel \noiselessN, having the same probability of errors $\{\lambda_{i\rightarrow j}|1 \le i \neq j \le M\}$ and $\{\lambda_{i \not\rightarrow i}|i=1,\ldots,M\}$.
\end{lemma}

\begin{IEEEproof}
We will  construct an ID code $\{(Q'_i, P_i) | i=1,\ldots, M\}$ with stochastic decoders for \noiselessN. Recall that $T_1,T_2,\cdots, T_N$ denote the distinct types for vectors in $\cAqn$.
For a type $T \in \cTqn$, $\cT_{T}^{(n)} \subset \cAqn$ denotes the \emph{type class} containing all $n$-length vectors with type $T$, and for any $\vx \in \cAqn$, $\cTxn$ denotes the set of all $n$-length vectors that share the same type as $\vx$.
For every $i\in [M]$, $j \in [N]$, we define
    \begin{align*}
        Q'_i(j) &\coloneqq Q_i(\cTn_{T_j})=\sum_{\vx\in \cTn_{T_j}}Q_i(\vx) ,\\
        P_i(1|j) &\coloneqq \frac{|\mathcal{D}_i \cap \cTn_{T_j}|}{|\cTn_{T_j}|}.
    \end{align*}
    The probability of errors for the new code are given by
    \begin{align*}
        \tilde{\lambda}_{i \rightarrow j} &\coloneqq \sum_{k \in [N]}Q'_i(k)P_j(1|k)\\
        &=\sum_{k \in [N]}Q'_i(k)\frac{|\mathcal{D}_j \cap \cTn_{T_k}|}{|\cTn_{T_k}|}\\
        &=\sum_{\bx \in \cAqn}Q_i(\bx)\frac{|\mathcal{D}_j \cap \cTxn|}{|\cTxn|} = \lambda_{i\rightarrow j},
    \end{align*}
    and
    \begin{align*}
        \tilde{\lambda}_{i \not\rightarrow i} &\coloneqq \sum_{k \in [N]}Q'_i(k)P_i(0|k)\\
        &=\sum_{k \in [N]}Q'_i(k)\frac{|\mathcal{D}_i^c \cap \cTn_{T_k}|}{|\cTn_{T_k}|}\\
        &=\sum_{\bx \in \cAqn}Q_i(\bx)\frac{|\mathcal{D}_i^c \cap \cTxn|}{|\cTxn|} = \lambda_{i \not\rightarrow i}.
    \end{align*}
    This proves the lemma.
\end{IEEEproof}

\subsection{From ID codes with stochastic decoders to ID codes with deterministic decoders for noiseless channels}

In this subsection, given an ID code with stochastic decoders for \noiselessN, we present a construction of deterministic decoders for \noiselessN, and provide a guarantee on the performance of the ID code comprising the same encoder and the new decoders. The result holds in general for any identification-feedback (IDF) code and any channel $P \in \cP(\cB|\cA)$, hence it is presented in this general form. In the proof of the soft converses, we will use it on the special case of ID codes (without feedback) for  \noiselessN. It was mentioned in~\cite{ahlswede1989identification}, \cite{ahlswede1989identification_f}, \cite{ahlswede1999identification} that using stochastic decoders does not increase the number of identifiable messages over a channel compared to deterministic decoders. This lemma supports the same fact.

 \begin{lemma}\label{lemma: st to det gen}
     Given an $(\numblocksfixed,M)$ IDF code $\{(\mbQ_{i,1}, \ldots,\mbQ_{i,\numblocksfixed}, P_i) | i=1,\ldots, M\}$ with stochastic encoder and decoders for a channel $P \in \cP(\cB|\cA)$, with error probabilities $\{\lambda_{i \rightarrow j}|1\le i \neq j \le M\}$ and $\{\lambda_{i \not\rightarrow i}|i=1,\ldots,M\}$, there exists $\cD_i \subset \cB^{\numblocksfixed}$, for $i=1,\ldots,M$, such that the $(l,M)$ IFD code $\{(\mbQ_{i,1}, \ldots,\mbQ_{i,\numblocksfixed},\cD_i)| i=1,\ldots,M\}$ with the same stochastic encoder 
 and deterministic decoders for $P \in \cP(\cB|\cA)$ has error probabilities bounded as
     \begin{align*}
         \tilde{\lambda}_{i \rightarrow j}  &\le \sqrt{\lambda_{i \rightarrow j}} \text{, for } 1 \le i \neq j \le M,\\
         \tilde{\lambda}_{i \not\rightarrow i} & \le \lambda_{i \not\rightarrow i} +  \sqrt{\lambda_2},\text{, for }i=1,\ldots,M,
     \end{align*}
     where $\lambda_2:= \max_{1 \le i\neq j \le M} \lambda_{i\rightarrow j}$.
 \end{lemma}

\begin{IEEEproof} 
    We fix any $0<\alpha<1$. Let us define, for each $i$, 
    \begin{align}
        \cD_i \coloneqq \{\vecy \in \cB^l | P_i(1|\vecy) > \alpha\}.\notag 
    \end{align}
    Let $\mbP_{\vecy}^{(i)}$ be the marginal distribution of $\mbP^{(i)}_{\vecx,\vecy}$ on $\vecy$.
    Then we have
    \begin{align}
         \lambda_{i \not\rightarrow i} &= \sum_{(y^{(1)}, \ldots, y^{(\numblocksfixed)}) \in \cB^{\numblocksfixed}} 
        \mbP^{(i)}_{\vecy}( y^{(1)},\ldots, y^{(\numblocksfixed)}) P_i(0| y^{(1)},\ldots, y^{(\numblocksfixed)}) \notag \\
        &=\sum_{(y^{(1)}, \ldots, y^{(\numblocksfixed)}) \in  \cD_i} 
        \mbP^{(i)}_{\vecy}( y^{(1)},\ldots, y^{(\numblocksfixed)}) P_i(0| y^{(1)},\ldots, y^{(\numblocksfixed)}) \notag \\
        &\hspace*{15mm}+\sum_{(y^{(1)}, \ldots, y^{(\numblocksfixed)}) \in  \cB^{l} \setminus \cD_i} 
        \mbP^{(i)}_{\vecy}( y^{(1)},\ldots, y^{(\numblocksfixed)}) P_i(0| y^{(1)},\ldots, y^{(\numblocksfixed)}) \notag \\
        &\ge 0+ \sum_{(y^{(1)}, \ldots, y^{(\numblocksfixed)}) \in  \cB^{l} \setminus \cD_i} 
        \mbP^{(i)}_{\vecy}( y^{(1)},\ldots, y^{(\numblocksfixed)}) (1-\alpha) \notag \\
        &=\sum_{(y^{(1)}, \ldots, y^{(\numblocksfixed)}) \in  \cB^{l} \setminus \cD_i} 
        \mbP^{(i)}_{\vecy}( y^{(1)},\ldots, y^{(\numblocksfixed)}) -\alpha \sum_{(y^{(1)}, \ldots, y^{(\numblocksfixed)}) \in  \cB^{l} \setminus \cD_i} 
        \mbP^{(i)}_{\vecy}( y^{(1)},\ldots, y^{(\numblocksfixed)}) \notag \\
        &\ge  \tilde{\lambda}_{i \not\rightarrow i}-\alpha \cdot 1. \notag 
    \end{align}
    Hence, we have
    \begin{align}
        \tilde{\lambda}_{i \not\rightarrow i} 
        &\le \lambda_{i \not\rightarrow i}+\alpha.\label{eq:st to det, i not i}
    \end{align}
    On the other hand,  we have, for $j \neq i$,
    \begin{align}
         \lambda_{i \rightarrow j} &= \sum_{(y^{(1)}, \ldots, y^{(\numblocksfixed)}) \in \cB^{\numblocksfixed}} 
        \mbP^{(i)}_{\vecy}( y^{(1)},\ldots, y^{(\numblocksfixed)}) P_j(1| y^{(1)},\ldots, y^{(\numblocksfixed)}) \notag \\
        &\ge \sum_{(y^{(1)}, \ldots, y^{(\numblocksfixed)}) \in \cD_j} 
        \mbP^{(i)}_{\vecy}( y^{(1)},\ldots, y^{(\numblocksfixed)}) P_j(1| y^{(1)},\ldots, y^{(\numblocksfixed)}) \notag \\
        &\ge \alpha \sum_{(y^{(1)}, \ldots, y^{(\numblocksfixed)}) \in \cD_j} 
        \mbP^{(i)}_{\vecy}( y^{(1)},\ldots, y^{(\numblocksfixed)}) = \alpha \tilde{\lambda}_{i \rightarrow j} .\notag 
    \end{align}
    Hence, we have
    \begin{align}
        \tilde{\lambda}_{i \rightarrow j} 
        &\le \lambda_{i \rightarrow j}/\alpha
    \end{align}
    The lemma follows by taking $\alpha= \sqrt{\lambda_2}$.
\end{IEEEproof}

\subsection{From non-uniform encoding distributions to uniform encoding distributions}

Let $\{(Q_i, D_i) | i=1,\ldots,M\}$ be an $(N,M)$ ID code with deterministic decoders designed for the channel \noiselessN. We now give a construction of a new code where the encoding distributions are uniform. The result is summarised in the following lemma.
The construction and the lemma are adapted from~\cite[Lemma $4$]{ahlswede1989identification}, which was used to prove the soft converse for second-order identification rates over DMCs.

\begin{lemma} \label{lemma: general to uniform}
    Given $\gamma \in (0,1)$, and an $(N,M)$ ID code with deterministic decoders $\{(Q_i,\cD_i)|i=1,\ldots,M\}$ for \noiselessN, with error probabilities $\{\lambda_{i \rightarrow j}|1\le i \neq j \le M\}$ and $\{\lambda_{i \not\rightarrow i}|i=1,\ldots,M\}$,  there exist sets $\mathcal{U}_i \subset [N]$ for $i=1,\ldots,M$ such that the ID code $\{(\cU_i,\cD_i)|i=1,\ldots,M\}$ with uniform encoding distributions has error probabilities satisfying
    \begin{align*}
        \tilde{\lambda}_{i \rightarrow j} &\le \lambda_{i \rightarrow j} \times \frac{(1+2\gamma)N^{\gamma}}{\gamma(1-N^{-\gamma})},\text{ for }1 \le i \neq j \le M,\\
        \tilde{\lambda}_{i \not\rightarrow i} &\le \lambda_{i \not\rightarrow i} \times \frac{(1+2\gamma)N^{\gamma}}{\gamma(1-N^{-\gamma})}, \text{ for } i=1,\ldots,M.
    \end{align*}
\end{lemma}
\begin{IEEEproof}
    We define $\kappa= \lceil \frac{1}{\gamma} \rceil+1$.
    For every $i \in [M]$ and $l\in[\kappa]$, we define
    \begin{align}
        \mathcal{B}(l,i) \coloneqq \{k \in [N]|N^{-\gamma l} <Q_i(k) \le N^{-\gamma(l-1)}\} \notag .
    \end{align}
    Then
    \begin{align}
        Q_i\left(\{[N] \setminus \cup_{l=1}^{\kappa}\mathcal{B}(l,i)\}\right) &\le 
        N\cdot N^{-\gamma \kappa} =  N^{-(\gamma \kappa-1)}
    \end{align}
    and hence
    \begin{align} \label{eq:1}
        Q_i\left( \cup_{l=1}^{\kappa}\mathcal{B}(l,i)\}\right) \ge (1- N^{-(\gamma \kappa-1)}).
    \end{align}
    Choose $l_i^* \coloneqq \text{argmax}_{l\in[\kappa]}  Q_i \left( \mathcal{B}(l,i)\right)$ and define
    \begin{align} \label{eq:2}
        \mathcal{U}_i \coloneqq \mathcal{B}(l_i^*,i ).
    \end{align}
    Combining \eqref{eq:1} and \eqref{eq:2}, we  can write
    \begin{align}\label{eq:3}
        Q_i(\mathcal{U}_i) \ge \frac{(1- N^{-(\gamma \kappa-1)})}{\kappa}.
    \end{align}
    Now let $Q_{\cU_i}(\cdot)$ denote the uniform distribution on $\cU_i$, i.e.,
    \begin{align*}
        Q_{\cU_i}(k)= \begin{cases}
            \frac{1}{|\mathcal{U}_i|},\text{ if } k \in \mathcal{U}_i\\
            0,\text{ otherwise}.
        \end{cases}
    \end{align*}
    Using \eqref{eq:3}, we can write, for $k \in \mathcal{U}_i$,
    \begin{align}
        Q_{\cU_i}(k) &= \frac{1}{|\mathcal{U}_i|} \notag \\
        &\le \frac{1}{|\mathcal{U}_i|} \times \frac{Q_i(\mathcal{U}_i)\kappa}{(1- N^{-(\gamma \kappa-1)})}. \label{eq:4}
    \end{align}
    For every $k,\hat{k} \in \mathcal{U}_i$, we have
    \begin{align*}
        Q_i(\hat{k}) &\le N^{-\gamma(l_i^*-1)}\\
        &=N^{\gamma} \cdot N^{-\gamma l_i^*}\\
        &\le N^{\gamma} Q_i(k).
    \end{align*}
    and summing both side over $\hat{k} \in \mathcal{U}_i$, we get
    \begin{align}
        Q_i(\mathcal{U}_i)=\sum_{\hat{k} \in \mathcal{U}_i}Q_i(\hat{k}) &\le \sum_{\hat{k} \in \mathcal{U}_i}N^{\gamma} Q_i(k) \notag \\
        &=|\mathcal{U}_i|N^{\gamma} Q_i(k).\label{eq:5}
    \end{align}
    Combining \eqref{eq:4} and \eqref{eq:5}, we can write, for every $k \in \mathcal{U}_i$,
    \begin{align}
        Q_{\cU_i}(k) &\le \frac{Q_i(\mathcal{U}_i)}{|\mathcal{U}_i|} \times \frac{\kappa}{(1- N^{-(\gamma \kappa-1)})} \notag \\
        &\le \frac{|\mathcal{U}_i|N^{\gamma} Q_i(k)}{|\mathcal{U}_i|} \times \frac{\kappa}{(1- N^{-(\gamma \kappa-1)})} \notag \\
        &=Q_i(k) \frac{\kappa N^{\gamma}}{(1- N^{-(\gamma \kappa-1)})}.
    \end{align}
Since $Q_{\cU_i}(k)=0$ for all $k\not\in \cU_i$, it follows that, for all $k\in [N]$,
    \begin{align}\label{eq:P_u_i<P_i}
        Q_{\cU_i}(k) &\le Q_i(k) \frac{\kappa N^{\gamma}}{(1- N^{-(\gamma \kappa-1)})}.
    \end{align}
We now bound the error probabilities for the ID code $\{(\cU_i,\cD_i)|i=1,\ldots,M\}$ with uniform encoding distributions.
    For $j \neq i$, we have
    \begin{align}
        \tilde{\lambda}_{i \rightarrow j} &= \sum_{k \in \cD_j}Q_{\cU_i}(k)\notag \\
        &\stackrel{}{\le} \frac{\kappa N^{\gamma}}{(1- N^{-(\gamma \kappa-1)})} \sum_{k \in \cD_j}Q_i(k) \notag \\
        &=\frac{\kappa N^{\gamma}}{(1- N^{-(\gamma \kappa-1)})} \times \lambda_{i \rightarrow j}\label{eq:uni i to j}.
    \end{align}   
    Similarly, for any $i$, we have
    \begin{align}
        \tilde{\lambda}_{i \not\rightarrow i} &= \sum_{k \in \cD_i^c}Q_{\cU_i}(k) \notag \\
        &\stackrel{}{\le} \frac{\kappa N^{\gamma}}{(1- N^{-(\gamma \kappa-1)})} \sum_{k \in \cD_i^c}Q_i(k) \notag \\
        &=\frac{\kappa N^{\gamma}}{(1- N^{-(\gamma \kappa-1)})} \times \lambda_{i \not\rightarrow i}. \label{eq:uni i not i}
    \end{align}
The lemma follows from
the bounds $\kappa \leq \frac{1+2\gamma}{\gamma}$ for the numerators, and $\kappa  > \frac{1+\gamma}{\gamma}$ for the denominators in \eqref{eq:uni i to j} and \eqref{eq:uni i not i}.  
\end{IEEEproof}

We will use Lemma~\ref{lemma: general to uniform} for $N$ denoting the number of types in $\cAqn$. 
By \eqref{eq:Nbounds}, we can then write $N \le 2n^{q-1}$ and $(1- N^{-\gamma})>1/2$ for large enough $n$.  
Hence the error probability bounds in Lemma~\ref{lemma: general to uniform} may be further upper bounded by
\begin{align}
        \tilde{\lambda}_{i \rightarrow j} &\le \lambda_{i \rightarrow j} \times \frac{12n^{ (q-1)\gamma}}{\gamma},\text{ for }1 \le i \neq j \le M, \label{eq:unif1}\\
        \tilde{\lambda}_{i \not\rightarrow i} &\le \lambda_{i \not\rightarrow i} \times \frac{12 n^{ (q-1)\gamma}}{\gamma}, \text{ for } i=1,\ldots,M. \label{eq:unif2}
\end{align}

\subsection{ID codes with decoding sets same as the support of encoding distributions}

Given an $(N,M)$ ID code $\{(\cA_i, \cD_i) | i=1,\ldots,M\}$ with deterministic decoders for \noiselessN,  we now give a construction of an ID code with deterministic decoders where the support of the uniform input distributions are same as the corresponding decoding regions, i.e., with $\cA_i=\cD_i$, $\forall i$.

\begin{lemma} \label{lemma:decoding equal support}
    Given an $(N,M)$ ID code  $\{(\cA_i,\cD_i)|i=1,\ldots,M\}$ with deterministic decoders and uniform encoder distributions  for \noiselessN, with error probabilities $\{\lambda_{i \rightarrow j}|1\le i \neq j \le M\}$ and $\{\lambda_{i \not\rightarrow i}|i=1,\ldots,M\}$, the code given by $\{(\cG_i,\cG_i)|i=1,\ldots,M\}$, where $\cG_i:=\cA_i\cap \cD_i$, 
    has error probabilities satisfying
    \begin{align}
       \tilde{\lambda}_{i \rightarrow j}
        \le \frac{\lambda_{i \rightarrow j}}{1-\lambda_{i \not\rightarrow i}}
    \end{align}
    for all $j \neq i$ and $\tilde{\lambda}_{i \not\rightarrow i}=0$ for all $i$.
\end{lemma}

\begin{IEEEproof}
Since the encoding distributions of the new code have the same supports as the corresponding decoding regions, clearly $\tilde{\lambda}_{i \not\rightarrow i}=0$ for all $i$.

For $j \neq i$, we have
\begin{align*}
    \tilde{\lambda}_{i \rightarrow j} 
    &=\frac{|\cG_i \cap \cG_j|}{|\cG_i|} \label{eq:n*ij}\\
    &=\frac{|\cG_i \cap \cG_j|}{|\cA_i|} \times \frac{|\cA_i|}{|\cG_i|}\\
    &\leq \frac{|\cA_i \cap \cD_j|}{|\cA_i|} \times \frac{|\cA_i|}{|\cG_i|}\\
    &= \lambda_{i \rightarrow j}\cdot \frac{1}{1-\lambda_{i \not\rightarrow i}}.
\end{align*}
 This proves the lemma.
\end{IEEEproof}

\subsection{ID codes with equal size support}
Given an ID code with decoding sets the same as the encoder supports, we now give a construction of a similar code where the size of all the supports is equal. We will later use the following lemma with $\lambda_1=0.$

\begin{lemma}\label{lemma:equal size support}
Given an $(N,M)$ ID code $\{(\cA_i, \cA_i)| i=1,\ldots,M\}$ with uniform encoder distributions and deterministic decoders for \noiselessN~having type-I/II error probabilities $\lambda_1,\lambda_2$, there exists an $\left(N,M'\right)$ ID code with equal size supports, having type-I and type-II error probabilities $\tilde{\lambda}_1\leq \lambda_1$, $\tilde{\lambda}_2\leq \lambda_2$, and $M'\ge \frac{M}{N}$.
\end{lemma}
\begin{IEEEproof} We prove the lemma by binning the support sizes and choosing the largest bin.
    For $k=1,\ldots,N$, we define the bins
\begin{align}
    \mathcal{N}(k) \coloneqq \{i \in \{1,\ldots,M\}\mid |\cA_i|=k\}.
\end{align}
By pigeon hole principle, there exists at least one set $\mathcal{N}(k^*)$ with 
$$M' \coloneqq |\cN(k^*)| \ge \frac{M}{N}.$$
We consider the new $\left(N,M'\right)$ ID code  $\{(\cA_i,\cA_i) | i \in \mathcal{N}(k^*) \}$. The Type-I and Type-II error probabilities of the new code clearly satisfy
\begin{align}
    \tilde{\lambda}_1 \le \lambda_1, \hspace{5mm} \text{ and }\hspace{5mm}
    \tilde{\lambda}_2 \le \lambda_2.
\end{align}
This proves the lemma.
\end{IEEEproof}

\section{Well Separated Set Systems}\label{sec:set_sysq}

In this section, we present some concepts and a proposition on set systems~\cite{jukna2011extremal}. This is used to prove the soft converse.

 For  $M,N \ge 1$ and $\Delta < \Gamma< N$, a $(\Gamma,\Delta)$ set system over $[N]$ is a collection $\mathcal{U}\coloneqq \{\mathcal{U}_1,\ldots,\mathcal{U}_{M}\}$ of distinct subsets of $[N]$ with  $|\mathcal{U}_i|=\Gamma$, $\forall i$, and maximum pairwise intersection size
$$
\Delta \coloneqq \max_{i\neq j} |\mathcal{U}_i\cap \mathcal{U}_j|.
$$
Note that if sets in a $(\Gamma,\Delta)$ set system over $[N]$ are used as encoding/decoding sets for an identification code for \noiselessN, then the sum probability of error $\lambda$ of the code is $\frac{\Delta}{\Gamma}$.

Suppose, $\Gamma > \frac{N}{2}$. Consider the collection of complement subsets $\tilde{\mathcal{U}}\coloneqq \{\mathcal{U}_1^c,\ldots,\mathcal{U}_{M}^c\}$ with set-sizes $\Gamma'=N-\Gamma$ and maximum size of pairwise intersections by $\Delta'$. 
Then 
\begin{align*}
    |\mathcal{U}_i^c\cap \mathcal{U}_j^c| &=|(\cU_i\cup \cU_j)^c|\\
    &= N - |\cU_i\cup \cU_j|\\
    &=N - (2\Gamma - |\cU_i\cap \cU_j|)\\
    &= N-2\Gamma +|\cU_i\cap \cU_j|
\end{align*}
Hence
\begin{align*}
    \Delta'
    &= N-2\Gamma +\Delta
\end{align*}
This gives us
\begin{align*}
    \frac{\Delta'}{\Gamma'}
    &= \frac{N-2\Gamma +\Delta}{N-\Gamma}\\
    &=1 - \frac{\Gamma - \Delta}{N-\Gamma}\\
    &\leq 1-\frac{\Gamma - \Delta}{\Gamma}\hspace*{3mm}\text {since } \Gamma> \frac{N}{2}\\
    &=\frac{\Delta}{\Gamma}
\end{align*}

Hence, for a given set system $\cU$ with $\Gamma>\frac{N}{2}$, the complementary set system $\tilde{\cU}$ has sets of size $\Gamma'\leq \frac{N}{2}$ and at least as small normalized intersection size $\frac{\Delta'}{\Gamma'}$.

\begin{proposition} \label{prop:well separated set system}
    Let $\alpha \in (0,1)$. Let $N\geq 1$ and $M> 1+\frac{N}{\alpha}$  be integers.  Let $\mathcal{U}\coloneqq \{\mathcal{U}_1,\ldots,\mathcal{U}_{M}\}$ be a $(\Gamma, \Delta)$ set system over $[N]$.
    Then
\begin{align}
    \intsec  \ge  (1-\alpha) h_2^{-1}\left( \frac{\log_2 M}{N}\right),
\end{align}
where  $h_2^{-1}(\cdot)$ is the inverse binary entropy function.
\end{proposition}

\begin{IEEEproof}
For $\Gamma>N/2$, we have seen above (before the proposition) that the normalized maximum pairwise-intersection ($\frac{\Delta'}{\Gamma'}$) of the collection of complement subsets is upper bounded by that of the given collection. Hence, without loss of generality, we assume that $1\le \Gamma\leq N/2$.

Let $\allones$ denote the $N$-length vector of all ones. For a subset $A\subseteq [N]$, the characteristic vector $\indicator{A}$ is an $N$-length vector with the $i$-th component $1$ if and only if $i\in A$.

Consider any $\epsilon \in (0,1)$. Associated with any subset $\cU \subseteq [N]$ with $|\cU|=N\epsilon$, we define an $N$-length vector
   \begin{align}
        \mathbf{v}_{\mathcal{U}} \coloneqq \left(\frac{\indicator{\mathcal{U}}-\epsilon \allones}{\sqrt{N\epsilon(1-\epsilon)}} \right)
    \end{align}
For any two subsets $\cU_1,\cU_2 \subseteq [N]$ with $|\cU_1|=|\cU_2|=N\epsilon $, we have
\begin{align}
	|\cU_1 \cap \cU_2| =\bigl\langle \indicator{\cU_1},\indicator{\cU_2}\bigr\rangle.
\end{align}
Therefore we can write
\begin{align}\label{eq:innr_product_set_vector}
	\langle \bv_{\cU_1},\bv_{\cU_2}\rangle&=\frac{1}{N\epsilon(1-\epsilon)} \left(\langle \indicator{\cU_1}-\epsilon\allones,\indicator{\cU_2}-\epsilon\allones \rangle\right) \notag \\
	&=\frac{1}{N\epsilon(1-\epsilon)}(\langle \indicator{\cU_1},\indicator{\cU_2} \rangle
        -\epsilon |\cU_1| \notag  \notag
	-\epsilon |\cU_2|+N\epsilon^2) \notag \\
	&=\frac{1}{N\epsilon(1-\epsilon)} \left(|\cU_1\cap \cU_2|-N\epsilon^2 \right).
\end{align}

Lemma~\ref{lemma:delta_n lower bound} below can be proved using Johnson's bound. However, we give an elementary proof here; the alternate proof using Johnson's bound is given in Appendix~\ref{sec:pf_by_johnson}.
\begin{lemma} \label{lemma:delta_n lower bound}
    Let $\alpha \in (0,1)$  and $N$ be fixed. 
    Consider a collection of subsets of $[N]$,  $\cU=\{\cU_1,\cdots,\cU_M\}$ with set sizes $N\epsilon$ and maximum pairwise intersection $N\delta$.
    If $M > 1+\frac{N}{\alpha}$ 
    then
    $\delta >(1-\alpha)\epsilon^2$.
\end{lemma}

\begin{IEEEproof}
We will prove by contradiction, that if $\delta \leq (1-\alpha)\epsilon^2$, then $M\leq 1+\frac{N}{\alpha}$.     By \eqref{eq:innr_product_set_vector}, for all  $i\neq j$, we have
    \begin{align}
        \langle \vv_{\cU_i},\vv_{\cU_j} \rangle \le -\frac{(\epsilon^2-\delta)}{\epsilon(1-\epsilon)}.
    \end{align}
    We can write
    \begin{align}
        0 \le \lVert\sum_{i=1}^{M} \vv_{\cU_i}\rVert_2^2 &= \sum_{i,j=1}^{M} \langle \vv_{\cU_i},\vv_{\cU_j} \rangle \notag \\
        &=\sum_{i}\lVert \vv_{\cU_i}\rVert^2+\sum_{i\neq j}\langle \vv_{\cU_i},\vv_{\cU_j} \rangle \notag \\
        &\le M-M(M-1)\frac{(\epsilon^2-\delta)}{\epsilon(1-\epsilon)}.\label{eq:lemma_4}
    \end{align}
Hence
    \begin{align}
        M &\le 1+\frac{\epsilon(1-\epsilon)}{(\epsilon^2-\delta)} \notag \\
    &\le 1+\frac{\epsilon}{(\epsilon^2-\delta)} \notag \\
    &\stackrel{(a)}{\le } 1+\frac{\epsilon}{\bigl(\epsilon^2-(1-\alpha)\epsilon^2\bigr)} \notag \\
    &\stackrel{}{= } 1+\frac{1}{\epsilon\alpha} \notag \\
    &\stackrel{(b)}{\le } 1+\frac{N}{\alpha} 
\end{align}
Here $(a)$ follows because $\delta \le (1-\alpha)\epsilon^2$, $(b)$ follows because $\epsilon \ge \frac{1}{N}$.
This is a contradiction, hence the lemma is proved.
\end{IEEEproof}

We now continue the proof of the proposition.  
As there are $\binom{N}{N\epsilon}$ number of distinct subsets of $[N]$ of size $N\epsilon$, we have
    \begin{align} \label{eq: epsilon>R_1}
        M &\le \binom{N}{N \epsilon} \notag \\
        &\stackrel{(a)}{\le} \frac{1}{\sqrt{N\pi \epsilon(1-\epsilon)}} 2^{Nh_2\left({\epsilon}\right)}
    \end{align}
    Here $(a)$ follows from~\cite[(17.41)]{cover1999elements}. As
    $
        1/N \le \epsilon \le 1/2,
    $
    we can upper bound the term 
    \begin{align}
        \frac{1}{\sqrt{N \pi \epsilon (1-\epsilon)}} \le \sqrt{\frac{2}{\pi}} \le 1.\notag 
    \end{align}
    Therefore
    \begin{align}
        M &\le  2^{Nh_2(\epsilon)},\notag 
    \end{align}
    i.e.
    \begin{align} \label{eq: epsilon lower bound}
        \epsilon \ge h_2^{-1}\left( \frac{\log_2 M}{N}\right).
    \end{align}
It is given in the proposition that $M > 1+\frac{N}{\alpha}$.
Therefore, by lemma~\ref{lemma:delta_n lower bound} and \eqref{eq: epsilon lower bound}
we have
\begin{align} \label{eq:Delta/Gamma1}
    \intsec=\frac{\delta}{\epsilon} &> (1-\alpha)\epsilon \notag \\
    &\ge (1-\alpha)h_2^{-1}\left( \frac{\log_2 M}{N}\right).
\end{align}
This proves the proposition.
\end{IEEEproof}

\section{Proof Of Theorem~\ref{thm:oneshot}}\label{sec:pf_thm_2A}
\subsection{Proof of Theorem~\ref{thm:oneshot}(\ref{thm:ach}) (achievability) }\label{app:pf_oneshot_ach}
Recall that $N$ is the number of types in $\cAqn$. First, for any $n$, we construct an ID code for \noiselessN. 
    
    Let us take
    \begin{align}
         \epsilon_n' &= \frac{n^{q-1}}{N}\left( \epsilon_n +\frac{1+\log N}{n^{q-1}} \right),\\
         \lambda_{2,n} &=\frac{4}{\log\left(\frac{1}{\epsilon_n'}\right)}, \label{eq:lambda_ach}\\
         \cZ&= [N].
    \end{align} 
    In Proposition~\ref{prop:ahlswede}. It fol
    lows from \eqref{eq:Nbounds} that $\epsilon_n'\rightarrow 0$ as $n\rightarrow \infty$, and hence for large enough $n$, $\epsilon_n'<\frac{1}{6}$. Also, for large enough $n$,
    \begin{align*}
    \lambda_{2,n} \log \left(\frac{1}{\epsilon_n'}-1\right) 
    & \geq \lambda_{2,n} \log \left(\frac{1}{2\epsilon_n'}\right)\\
    & = \lambda_{2,n} \times \left(\log \left(\frac{1}{\epsilon_n'} \right)-1\right) \\
    & \geq  \frac{4}{\log\left(\frac{1}{\epsilon_n'}\right)} \times \frac{1}{2}\log  \left(\frac{1}{\epsilon_n'} \right)\\
    & = 2
    \end{align*}
    Hence $\epsilon_n'$ and $\lambda_{2,n}$ satisfy the conditions in  Proposition~\ref{prop:ahlswede}, which guarantees that there exist
    \begin{align*}
        M &\ge 2^{\epsilon_n' N-1-\log N}\\
        &= 2^{\bigg\{  \frac{n^{q-1}}{N}\left( \epsilon_n +\frac{1+\log N}{n^{q-1}} \right)  \cdot N -1 -\log N \bigg\}}\\
        &=2^{\big\{ \epsilon_n n^{q-1} \big\}}
    \end{align*}
    number of distinct subsets $\{\cU_1, \ldots, \cU_M\}$ of $[N]$ such that
    \begin{align*}
        |\cU_i|&=\epsilon_n' N, \,\,\forall i,\\
        |\cU_i \cap \cU_j| &\le \lambda_{2,n} \epsilon_n' N, \,\,\forall i \neq j. 
    \end{align*}
This collection gives an ID code $\{(\cU_i,\cU_i)|i=1,\ldots,M)\}$ for \noiselessN~with probability of errors $\lambda_1=0$, $\lambda_2=\lambda_{2,n}$.

Now, since permutations preserve the type of a vector, the above code can be used to construct an ID code for \bupcnq\!\!. Suppose $\vx_1,\vx_2,\cdots,\vx_N$ are a set of representatives from the $N$ distinct type classes $\cT_1,\cT_2,\cdots, \cT_N$ respectively in $\cAqn$. Then we define the code $\{(Q_i,\cD_i)|i=1,\ldots,M)\}$ where $Q_i$ is the uniform distribution over $\{\vx_j|j\in \cU_i\}$, and 
\begin{align}
    \cD_i = \cup_{j\in \cU_i} \cT_j.\notag 
\end{align}
Since permutations do not change the type of a transmitted vector, the probabilities of error for this code are also $\lambda_1=0$, $\lambda_2=\lambda_{2,n}$. Since $\lambda_{2,n} \rightarrow 0$, this proves Theorem~\ref{thm:oneshot}(\ref{thm:ach}).

\subsection{Proof of Theorem~\ref{thm:oneshot}(\ref{thm:soft_converse}) (soft converse)}\label{sec:pf thm soft converse oneshot}

Consider a given sequence of $(n_i,2^{n_i^{q-1}R},\lambda_{1,n_i},\lambda_{2,n_i})$ ID codes for  \bupcq  with  $n_i \rightarrow \infty$ and probability of errors $\lambda_{1,n_i},\lambda_{2,n_i} < {n_i}^{-\errorpower}$ for all $i$.
Through a series of modification steps, we will prove the existence of a sequence of ID codes for \noiselessN~with uniform encoding distributions, decoding regions the same as the support of the corresponding encoding distributions, and vanishing error probability. We will then show using Proposition~\ref{prop:well separated set system} that for such a sequence of codes to exist, the rate $R$ must be $0$. 

{\bf Step 0:} We start with any $(n,2^{n^{q-1}R},\lambda_{1,n},\lambda_{2,n})$ ID code in the sequence for $n=n_i$ and $\lambda_{1,n}, \lambda_{2,n} \le n^{-\errorpower}$. We assume large enough $n$.

{\bf Step 1:} From the given code in Step 0, using Lemma~\ref{lemma: permutation to noiseless}, we construct an $(N, 2^{n^{q-1}R},n^{-\errorpower},n^{-\errorpower})$ ID code for \noiselessN~with stochastic decoders. Here $N=\left|\cTqn\right|$.

{\bf Step 2:} From the given code in Step 1, we use Lemma~\ref{lemma: st to det gen} to construct an 
$(N,2^{Rn^{q-1}},2n^{-\errorpower/2},n^{-\errorpower/2})$ ID code with deterministic decoders for \noiselessN.

{\bf Step 3:} From the given code in Step 2, we use Lemma~\ref{lemma: general to uniform} (see \eqref{eq:unif1} and \eqref{eq:unif2}) with $\gamma= \frac{\errorpower}{4(q-1)}$, to construct an  $\left(N,2^{Rn^{q-1}},2Kn^{-\errorpower/4},Kn^{-\errorpower/4}\right)$ ID code for \noiselessN~with deterministic decoders and uniform encoding distributions. Here $K$ is some constant which depends on $q$ and $\errorpower$.

{\bf Step 4:} From the given code in Step 3, we use  Lemma~\ref{lemma:decoding equal support} to construct an  $\left(N,2^{Rn^{q-1}},0,2Kn^{-\errorpower/4}\right)$ ID Code for \noiselessN~with deterministic decoders, and decoding regions same as the support of corresponding uniform encoder distributions. We assume $n$ to be large enough s.t. $Kn^{-\errorpower/4} < 1/2$.

 {\bf Step 5:}  From the given code in Step 4, we use  Lemma~\ref{lemma:equal size support} to construct an   $\left(N,2^{R_n n^{q-1}},0,2Kn^{-\errorpower/4}\right)$ ID code for \noiselessN~with deterministic decoders and equal size supports. Here $R_n $ satisfies  $R_n\ge R-\frac{\log N}{n^{q-1}} $.

{\bf Final argument:}
The sequence of codes obtained in Step 5 from the original sequence of codes for \bupcq gives a sequence of set systems with
$M_{i}=2^{{n_i^{q-1}}R_{n_i}}$, and $\frac{\Delta_{i}}{\Gamma_{i}}\le 2K{n_i}^{-\errorpower/4}$. Hence $\liminf_{i \rightarrow \infty} \frac{\Delta_{n_i}}{\Gamma_{n_i}}=0$. 
On the other hand, by using the bounds on $N$ in \eqref{eq:Nbounds}, for an $n_i$, we have
\begin{align}\label{eq:Nbounds2}
N_i \le (n_i+q-1)^{(q-1)};
\end{align}
By Proposition~\ref{prop:well separated set system}, for $\alpha=1/2$, we have
\begin{align*}
    \liminf_{i \rightarrow \infty} \frac{\Delta_{n_i}}{\Gamma_{n_i}}& \geq \frac{1}{2} \liminf_{i \rightarrow \infty} h_2^{-1}\left( \frac{\log (2^{R_{n_i} {n_i}^{(q-1)}})}{N_i}\right)\\
    &\stackrel{}{\ge}  \frac{1}{2} \liminf_{i \rightarrow \infty} h_2^{-1}\left( \frac{{n_i}^{(q-1)}R_{n_i}}{({n_i}+q-1)^{(q-1)}}\right)
    \\
    &=\frac{1}{2} \liminf_{i \rightarrow \infty} h_2^{-1}\left(R_{n_i} \bigg[\frac{{n_i}}{({n_i}+q-1)}\bigg]^{q-1}\right)\\
    &=\frac{1}{2} \liminf_{i \rightarrow \infty} h_2^{-1}\left(R_{n_i} \bigg[\frac{1}{(1+q/{n_i}-1/{n_i})}\bigg]^{q-1}\right)\\
    &\geq \frac{1}{2} \liminf_{i \rightarrow \infty} h_2^{-1}\bigg(\bigg[R-\frac{(q-1)\log ({n_i}+q-1)}{{n_i}^{q-1}}\bigg] \\
    & \hspace{20mm}\bigg[\frac{1}{(1+q/{n_i}-1/{n_i})}\bigg]^{q-1} \bigg)\\
    &= \frac{1}{2}h_2^{-1}(R) >0 \text{ if } R >0.
\end{align*}
This gives a contradiction if $R>0$. This completes the proof of Theorem~\ref{thm:oneshot}(\ref{thm:soft_converse}).

\subsection{Proof of Theorem~\ref{thm:oneshot}(\ref{thm:strong_converse}) (strong converse)}\label{sec:strong converse oneshot proof}
Towards proving Theorem~\ref{thm:oneshot}(\ref{thm:strong_converse}), we use a result on the approximation of probability distributions with distributions having certain ``resolution'' \cite{han1993approximation,bocherer2016optimal}.
The variational distance between the two random variables $X,Y$ taking values from a set $\cZ$ and having distributions $P_X$ and $P_Y$ respectively,  is defined as
    \begin{align}
        d(X,Y) &\coloneqq \sum_{z \in \cZ} \left|P_X(z)-P_Y(z)\right|.
    \end{align}
We shall use $d(P_X,P_Y)$ and $d(X,Y)$ interchangeably.
For any random variable $Y$, the following lemma guarantees the existence of an approximating distribution of a certain resolution.
The lemma is a special case of~\cite[Proposition $2$]{bocherer2016optimal} and \cite[Lemma 2.1.1]{koga2002information}.

\begin{lemma}\label{lemma:approx output dist}
	For any positive integer $N$, and any random variable $Y$ taking values in $[N]$, and arbitrary constants $\alpha>0$,
	 there exists a mapping $\phi:[N^{\alpha}] \rightarrow [N]$ that satisfies  
	\begin{align}
		d(Y,&\phi(U_{N^{\alpha}})) \le N^{-(\alpha-1)}
  \label{eq:approx1}
	\end{align}
	where 
 $U_{N^{\alpha}}$ is the uniform random variable on $[N^{\alpha}]$. 
\end{lemma}

\begin{remark}
    Note that the right-hand side of \eqref{eq:approx1} vanishes as $N\rightarrow \infty$ if only if $\alpha>1$.
\end{remark}

We now proceed to prove Theorem~\ref{thm:oneshot}(\ref{thm:strong_converse}).
Consider a given sequence of $(n_i,2^{\Rni n_i^{q-1}},\lambda_{1,i}, \lambda_{2,i})$ ID codes for  \bupcq  with  $n_i \rightarrow \infty$ and $\Rni \rightarrow \infty$ as $i \rightarrow \infty$ (i.e. $n_i \rightarrow \infty$).
Recall that $N_i$ is the number of type classes in $\cA_q^{n_i}$.
For each $i$,  using Lemma~\ref{lemma: permutation to noiseless}, we construct an $(N_i, 2^{\Rni n_i^{q-1}},\lambda_{1,i}, \lambda_{2,i})$ ID code $\{(Q_j^{(i)},P_j^{(i)})|j=1,\ldots,2^{\Rni n_i^{(q-1)}}\}$ for \noiselessNi with stochastic decoders.  We will now show that for such a sequence of codes, $\liminf_{i \rightarrow \infty} (\lambda_{1,i}+\lambda_{2,i}) \ge 1$.

    Fix an $i$. The total variational distance between any two encoding distributions $Q_j^{(i)}, Q_k^{(i)}$ (for messages $j$ and $k$ respectively) is related to the error probability  $\lambda_i$ as follows. For all $j \neq k$, we have
    \begin{align}
        d(Q_j^{(i)},Q_k^{(i)}) & = \sum_{x \in [N_i]} |Q_j^{(i)}(x)-Q_k^{(i)}(x)| \notag \\
	&\ge  \sum_{x \in [N_i]} P_j^{(i)}(1|x) |Q_j^{(i)}(x)-Q_k^{(i)}(x)| \notag \\
        &\ge  \bigg(\sum_{x \in [N_i]} P_j^{(i)}(1|x)Q_j^{(i)}(x) 
        -  \sum_{x \in [N_i]} P_j^{(i)}(1|x)Q_k^{(i)}(x) \bigg)  \notag \\
	&=1-\lambda_{i,j\not\rightarrow j}-\lambda_{i,k\rightarrow j}\\
	&\ge  1-\lambda_{1,i} - \lambda_{2,i}, \label{eq:rmk1_1}
    \end{align}
    i.e.,
    \begin{align} \label{eq: lambda_n and TV distance}
        \lambda_{1,i}+\lambda_{2,i} \ge 1-d(Q_j^{(i)},Q_k^{(i)}), \hspace{.2cm} \forall j \neq k, 
    \end{align}

We define  
\begin{align}
    \Rni'=\frac{\Rni}{4(q-1)\log n_i } \label{eq:\Rni'}
\end{align}
and  $\alpha_{i}=1+\Rni'$. 
By Lemma~\ref{lemma:approx output dist}, for every $j \in [2^{\Rni n_i^{q-1}}]$, there exists a mapping $\phi_j^{(i)}:[N_i^{\alpha_{i}}] \rightarrow [N_i]$  such that the random variable $\tilde{Y}_i(j) \coloneqq \phi_j^{(i)}(U_{N_i^{\alpha_{i}}})$ satisfies 
\begin{align}
	d\left(Q_j^{(i)},Q_{\tilde{Y}_i(j)}\right) \le \delta_i, \notag 
\end{align}
where 
\begin{align}
    \delta_i &=   N_i^{-(\alpha_{i}-1)}
    =N_i^{-\Rni'}=2^{-\Rni' \log N_i}& \notag \\
    &\stackrel{(a)}{\le} 
    2^{-\Rni' \left(1-\frac{\log (q-1)!}{(q-1)\log n_i} \right) (q-1)\log n_i  }& \notag \\
    &\stackrel{}{\le} 2^{-\frac{1}{2}\times \Rni' (q-1)\log n_i}&\text{(for large enough $i$)} \notag \\ 
    &=2^{-\frac{\Rni}{8}},\label{eq:delta_bound_strong_oneshot}
\end{align}
here $(a)$ follows from the lower bound in \eqref{eq:Nbounds}. Since $\Rni \rightarrow \infty$, we have  $\delta_{i}\rightarrow 0$ as $i \rightarrow \infty$.

 The number of distinct $N_i^{\alpha_{i}}$-type distributions on $[N_i]$ is upper bounded by using \eqref{eq:Nbounds} (where $n$ is replaced with $N_i^{\alpha_{i}}$ and $q$ is replaced with $N_i$) as
 \begin{align}
     \binom{N_i^{\alpha_{i}}+N_i-1}{N_i-1} 
     &\le \frac{N_i^{\alpha_{i}(N_i-1)}}{(N_i-1)!}\left(1+\frac{N_i-1}{N_i^{\alpha_{i}}}\right)^{N_i-1}\notag&\\
     &\le \frac{N_i \cdot N_i^{\alpha_{i}(N_i-1)}}{N_i!}\cdot 2^{N_i}\notag & \\
     &\stackrel{}{\le}(2e)^{N_i}  \times 
     \frac{N_i^{[\alpha_{i}(N_i-1)+1]}}{N_i^{N_i}}\notag & \text{ (by Stirling's approximation)}\notag \\
     &=N_i^{\{ \alpha_{i} N_i -\alpha_{i} +1 -N_i+N_i\log_{N_i} (2e) \}}\notag& \\
     &=N_i^{N_i\left(\alpha_{i} -1 +\frac{\log (2e)}{\log  N_i}-\frac{\alpha_{i}-1}{N_i} \right)}& \notag \\
     &=N_i^{N_i\left((\alpha_{i} -1)\left(1-\frac{1}{N_i} \right) +\frac{\log (2e)}{\log  N_i} \right)}
     &\notag \\
     &\stackrel{(a)}{\le} N_i^{N_i\left(\Rni'+\frac{\log(2e) }{(q-1)\left( 1-\frac{\log (q-1)!}{(q-1)\log n_i}\right)\log n_i}\right)} &\notag \\
     &\stackrel{}{\le} N_i ^{N_i \left(\Rni'+\frac{2\log(2e) }{(q-1)\log n_i}\right) } &\text{ (for large enough $i$)}\notag \\
     &\stackrel{(b)}{=} N_i ^{N_i \left(\Rni'+\frac{8\Rni'\log(2e) }{\Rni}\right) }&\notag \\
     &\stackrel{}{=}N_i ^{N_i \left(\Rni' \left(1+\frac{8\log(2e) }{\Rni}\right)\right) }&\notag \\
     &\stackrel{}{\le}N_i ^{2 N_i \Rni' } &\text{ (for large enough $i$)}\notag \\
     &=2^{2\Rni' N_i\log N_i}&\notag \\
    &\stackrel{(c)}{\le} 2^{2\Rni' n_i^{q-1} \left(1+\frac{q-1}{n_i} \right)^{q-1}  \log \left(  n_i^{q-1} \left(1+\frac{q-1}{n_i} \right)^{q-1}\right)}&\notag \\
    &=2^{2\Rni' \left(1+\frac{q-1}{n} \right)^{q-1} n_i^{q-1} (q-1)\left(1+\frac{\log \left( 1+\frac{q-1}{n_i}\right)}{\log n_i} \right)\log n_i}&\notag \\
    & \stackrel{(d)}{\le} 2^{\frac{R_i}{2} \left(1+\frac{q-1}{n_i}\right)^{q-1} \left(1+\frac{1}{\log n_i} \right)} & \text{ (for large enough $i$)}\\
    &< 2^{R_i n_i^{q-1}}&\text{ (for large enough $i$)}.\label{eq:type_size_strong}
\end{align}
Here $(a)$ follows because of \eqref{eq:Nbounds}, by which $\log N_i \ge (q-1)\left( 1-\frac{\log (q-1)!}{(q-1)\log n_i}\right)\log n_i$, $(b)$ follows from \eqref{eq:\Rni'}, and $(c)$ follows from \eqref{eq:Nbounds}, $(d)$ follows by using $R_i=4\Rni'  (q-1)\log n_i$, and $\log \left( 1+\frac{q-1}{n_i}\right)\leq 1$ for large enough $n_i$.

We have shown that for large enough $i$, the number of distinct $N_i^{\alpha_{i}}$-type distributions on $[N_i]$ is less than $2^{R_in_i^{q-1}}$.
Hence, $\exists I$ such that for each $i>I$, there exist at least one pair $(j,k)$ with $j\neq k$ and $Q_{\tilde{Y}_i(j)}=Q_{\tilde{Y}_i(k)}$ and  we have
\begin{align}
	d(Q_j^{(i)},Q_k^{(i)})&\le d\left(Q_j^{(i)},Q_{\tilde{Y}_i(j)}\right)+d\left(Q_k^{(i)},Q_{\tilde{Y}_i(j)}\right)\notag \\
	&= d\left(Q_j^{(i)},Q_{\tilde{Y}_i(j)}\right)+ d\left(Q_k^{(i)},Q_{\tilde{Y}_i(k)}\right)\notag \\
	&\le 2 \delta_i. \label{eq:TV_lower_bound}
\end{align}
By combining \eqref{eq: lambda_n and TV distance} and \eqref{eq:TV_lower_bound} we have
\begin{align}\label{eq:rmk1_2}
    \lambda_{1,i}+\lambda_{2,i} \ge 1-2\delta_{i}.
\end{align}
Since $\delta_{i}\rightarrow 0$ as $i \rightarrow \infty$, it follows that
\begin{align}\label{eq:rmk1_3}
    \liminf_{i \rightarrow \infty} (\lambda_{1,i}+\lambda_{2,i}) \geq 1.
\end{align}
This completes the proof of Theorem~\ref{thm:oneshot}(\ref{thm:strong_converse}).
\begin{remark}
    The bound on the number of messages in the strong converse is weaker than that in the weak converse, since $R_i\rightarrow \infty$. This weakness comes due to the tension between the requirement of ensuring $\delta_i\rightarrow 0$ in \eqref{eq:delta_bound_strong_oneshot} and also upper bounding the number of distinct $N_i^{\alpha_i}$-type distribution on $[N_i]$ to the number of messages in \eqref{eq:type_size_strong}. We can use a different scaling of $R_i$ to define $R'_i=R_i/c_i$ for some $c_i$ (we have used $c_i=4(q-1)\log n_i$ in the proof). But it can be checked that, to ensure that $\delta_i\rightarrow 0$, the bound on the message size will still be of the form $2^{R''_i n^{q-1}}$ where $R''_i\rightarrow \infty$. 
\end{remark}

\section{Proof Of Theorem~\ref{thm:qary_multishot}} \label{sec:multishot}
\subsection{Proof of Theorem~\ref{thm:qary_multishot}(\ref{thm:achv_multishot}) (achievability)}
\label{app:pf_multishot_ach}
The proof of Theorem~\ref{thm:qary_multishot}(\ref{thm:achv_multishot}) follows by similar arguments as the proof of Theorem~\ref{thm:oneshot}(\ref{thm:ach}).
We first construct an ID code $\{(\cU_i,\cU_i)|i=1,\ldots,M)\}$  for $\nll$ similarly as in the proof of Theorem~\ref{thm:oneshot}(\ref{thm:ach}) by replacing $N$ by $N^{\numblocksfixed}$. This code has  probabilities of errors $\lambda_{1,i}=0$, $\lambda_{2,i} \rightarrow 0$.

Now suppose $\vx_1,\vx_2,\cdots,\vx_N$ are a set of representatives from the $N$ distinct type classes $\cT_1,\cT_2,\cdots, \cT_N$ respectively in $\cAqn$.  
Note that there exists a bijection $\phi: [N]^{\numblocksfixed} \rightarrow [N^{\numblocksfixed}]$.
Let us denote $\phi(j_1, \ldots,j_{\numblocksfixed})=j$ where $j_1,\ldots,j_l \in [N]$. 
We define an $(l,M)$ ID code $\{(Q_i,\cD_i)|i=1,\ldots,M)\}$ for \bupcnq, where $Q_i$ is the uniform distribution over $\{(\vx_{j_1}, \ldots,\vx_{j_{\numblocksfixed}})| j_1, \ldots, j_l \in [N], \phi(j_1, \ldots, j_{\numblocksfixed})\in \cU_i\}$, and 
\begin{align}
    \cD_i = \cup \{\cT_{j_1} \times \cdots \times \cT_{j_{\numblocksfixed}}\},\notag 
\end{align}
where the union is over $\{(j_1,\ldots,j_l) \in [N]^l|\phi(j_1, \ldots, j_{\numblocksfixed}) \in \cU_i\}$.
Since permutations do not change the type of a transmitted vector, the probabilities of error for this code are also $\lambda_{1,i}=0$,  $\lambda_{2,i} \rightarrow 0$. This completes the proof.
\subsection{Proof  of Theorem~\ref{thm:qary_multishot}(\ref{thm:soft_converse_multishot_i}) (soft converse 1 )}\label{sec:proof soft 1 multishot}
Consider a given sequence of $(n_i,l_i,2^{n_i^{l_i(q-1)}R},\lambda_{1,n_i},\lambda_{2,n_i})$ ID codes for  \bupcq  with  $n_i \rightarrow \infty$ and probability of errors $\lambda_{1,n_i},\lambda_{2,n_i} < {n_i}^{-\errorpower}$ for all $i$, and $l_i < n_i^{\omega}$ where $\omega < \min{\{1,\errorpower/4\}}$. By fixing an $i$, we start with an $(n,l,2^{n^{l(q-1)}R},\lambda_{1,n},\lambda_{2,n})$ ID code in the sequence for $n=n_i,l<n^{\omega}$, where $\omega < \min{\{1,\errorpower/4\}}$, and $\lambda_{1,n}, \lambda_{2,n} \le n^{-\errorpower}$. We assume large enough $n$, and follow the steps 1-5 as in the proof of Theorem~\ref{thm:oneshot}(\ref{thm:soft_converse}), with some modifications. In the following, we outline these steps with suitable modifications, where needed. All the steps mentioned below refer to the steps in the proof of Theorem~\ref{thm:oneshot}(\ref{thm:soft_converse}) (Sec.~\ref{sec:pf thm soft converse oneshot}).

{\bf Modified Step 1:}
For step 1, we need a modified multishot version of Lemma~\ref{lemma: permutation to noiseless}.
\begin{lemma}[Multishot version of Lemma~\ref{lemma: permutation to noiseless}]\label{lemma: permutation to noiseless_multishot}
    Given an $(n,l,M)$ ID code with deterministic decoders $\{(Q_i, \mathcal{D}_i) | i=1,\ldots, M\}$ for \bupcnq\!,
     there exists an $(N^l,M)$ ID code $\{Q'_i,P_i| i=1,\ldots, M\}$ with stochastic decoders for the noiseless channel $\nll$, having the same probability of errors $\{\lambda_{i\rightarrow j}|1 \le i \neq j \le M\}$ and $\{\lambda_{i \not\rightarrow i}|i=1,\ldots,M\}$.
\end{lemma}
The proof of Lemma~\ref{lemma: permutation to noiseless_multishot} can be found in Appendix~\ref{app:proof permutation to noiseless_multishot}.
From the given code in Step 0, we use Lemma~\ref{lemma: permutation to noiseless_multishot} to construct an $(N^{l}, 2^{n^{l(q-1)}R},n^{-\errorpower},n^{-\errorpower})$ ID code for $\nll$ with stochastic decoders. Here $N=\left|\cTqn\right|$.

{\bf Step 2:} From the given code in Step 1, by using Lemma~\ref{lemma: st to det gen}  where $N$ is replaced with $N^{\numblocksfixed}$, this step gives  an 
$(N^l,2^{Rn^{l(q-1)}},2n^{-\errorpower/2},n^{-\errorpower/2})$ ID code with deterministic decoders for $\nll$.

{\bf Modified Step 3:} From the given code in Step 2, we use Lemma~\ref{lemma: general to uniform} where $N$ is replaced with $N^{\numblocksfixed}$ and set $\gamma= \frac{\errorpower}{4l(q-1)}$ to construct an  $\left(N^l,2^{Rn^{l(q-1)}},\frac{8(2lq+\errorpower)}{\errorpower}n^{-\errorpower/4},\frac{4(2lq+\errorpower)}{\errorpower}n^{-\errorpower/4}\right)$ 
ID code for $\nll$ with deterministic decoders and uniform encoding distributions. In addition to replacing $N$ with $N^l$, a key modification here is the different value of $\gamma$.

{\bf Step 4:} From the given code in Step 3, we use  Lemma~\ref{lemma:decoding equal support} where $N$ is replaced with $N^{\numblocksfixed}$  to construct an  $\left(N^l,2^{Rn^{l(q-1)}},0,\frac{8(2lq+\errorpower)}{\errorpower}n^{-\errorpower/4} \right)$ ID Code for $\nll$ with deterministic decoders, and decoding regions same as the support of corresponding uniform encoder distributions. Here we use the fact that for large enough $n$, $\frac{4(2lq+\errorpower)}{\errorpower}n^{-\errorpower/4} < 1/2$, if $l < n^{\omega}$ for some $\omega<\errorpower/4$ 

{\bf Step 5:}  From the given code in Step 4, we use  Lemma~\ref{lemma:equal size support} where $N$ is replaced with $N^{\numblocksfixed}$  to construct an   $\left(N^l,2^{R_n n^{l(q-1)}},0,\frac{8(2lq+\errorpower)}{\errorpower}n^{-\errorpower/4}\right)$ ID code for $\nll$ with deterministic decoders and equal size supports. Here $R_n $ satisfies  $R_n\ge R-\frac{l\log N}{n^{l(q-1)}} $.

{\bf Final argument:}
The sequence of $\left(N_i^{l_i},2^{R_{n_i} n_i^{l_i(q-1)}},0,\frac{8(2l_iq+\errorpower)}{\errorpower}n_i^{-\errorpower/4}\right)$ ID codes for $\text{NL}_{[N_i^{\numblocksfixed_i}]}$ obtained in Step 5 from the original sequence of codes for \bupcq gives a sequence of set systems with
$M_{i}=2^{{n_i^{l_i(q-1)}}R_{n_i}}$, and $\frac{\Delta_{i}}{\Gamma_{i}}\le \frac{8(2l_iq+\errorpower)}{\errorpower}n_i^{-\errorpower/4}$. Hence $\liminf_{i \rightarrow \infty} \frac{\Delta_{i}}{\Gamma_{i}}=0$ as $l_i<n_i^\omega$ for some $\omega<\mu/4$.
On the other hand,
by Proposition~\ref{prop:well separated set system}, for $\alpha=1/2$, we have
\begin{align*}
    \liminf_{i \rightarrow \infty} \frac{\Delta_{i}}{\Gamma_{i}}& \stackrel{}{\geq} \frac{1}{2} \liminf_{i \rightarrow \infty} h_2^{-1}\left( \frac{\log (2^{R_{n_i} {n_i}^{l_i(q-1)}})}{N_i^{l_i}}\right)
    \stackrel{(a)}{ \ge } \frac{1}{2}h_2^{-1}(R) >0 \text{, if } R >0.
\end{align*}
Inequality $(a)$ follows because $l_i =o(n_i^{\omega})$ where $\omega < 1$.
This gives a contradiction if $R>0$. This completes the proof of Theorem~\ref{thm:qary_multishot}(\ref{thm:soft_converse_multishot_i}).

\subsection{Proof  of Theorem~\ref{thm:qary_multishot}(\ref{thm:soft_converse_multishot_ii}) (soft converse 2)}\label{sec:proof soft 2 multishot}
Consider a given sequence of $(n_i,l_i,2^{n_i^{l_i(q-1)}R},\lambda_{1,n_i},\lambda_{2,n_i})$ ID codes for  \bupcq  with  $n_i \rightarrow \infty$, $l_i=o(n_i)$, and probability of errors $\lambda_{1,n_i},\lambda_{2,n_i} < {n_i}^{-l_i\errorpower}$ for all $i$. 
We start with an $(n,l,2^{n^{l(q-1)}R},\lambda_{1,n},\lambda_{2,n})$ ID code in the sequence for $n=n_i, l=o(n)$, and $\lambda_{1,n}, \lambda_{2,n} \le n^{-l\errorpower}$. We assume large enough $n$.

The proof of Theorem~\ref{thm:qary_multishot}(\ref{thm:soft_converse_multishot_ii}) closely follows that of Theorem~\ref{thm:qary_multishot}(\ref{thm:soft_converse_multishot_i}), with identical steps apart from certain modifications. We will outline these modifications without repeating the full proof.

{\bf Modified Step 3:} From the given code in Step 2, we use Lemma~\ref{lemma: general to uniform} where $N$ is replaced with $N^{\numblocksfixed}$, and set $\gamma= \frac{\errorpower}{4(q-1)}$, to construct an  $\left(N^l,2^{Rn^{l(q-1)}},2K n^{-l\errorpower/4},Kn^{-l\errorpower/4}\right)$ 
ID code for $\nll$ with deterministic decoders and uniform encoding distributions. Here $K$ is some constant which depends on $q$ and $\errorpower$.

{\bf Final argument:}
The sequence of $\left(N_i^{l_i},2^{R_{n_i} n_i^{l_i(q-1)}},0,2Kn_i^{-l_i\errorpower/4}\right)$ ID codes for $\text{NL}_{[N_i^{l_i}]}$ obtained in Step 5 from the original sequence of codes for \bupcq gives a sequence of set systems with
$M_{i}=2^{{n_i^{l_i(q-1)}}R_{n_i}}$, and $\frac{\Delta_{i}}{\Gamma_{i}}\le 2K n_i^{-l\errorpower/4}$. Hence $\liminf_{i \rightarrow \infty} \frac{\Delta_{i}}{\Gamma_{i}}=0$.
On the other hand, by Proposition~\ref{prop:well separated set system}, for $\alpha=1/2$, we have
\begin{align*}
    \liminf_{i \rightarrow \infty} \frac{\Delta_{i}}{\Gamma_{i}}& \stackrel{}{\geq} \frac{1}{2} \liminf_{i \rightarrow \infty} h_2^{-1}\left( \frac{\log (2^{R_{n_i} {n_i}^{l_i(q-1)}})}{N_i^{l_i}}\right)
    \stackrel{(a)}{ \ge } \frac{1}{2}h_2^{-1}(R) >0 \text{ if } R >0.
\end{align*}
The inequality $(a)$ follows because $l_i=o(n_i)$.
This gives a contradiction if $R>0$. This completes the proof of Theorem~\ref{thm:qary_multishot}(\ref{thm:soft_converse_multishot_ii}).

\subsection{Proof  of Theorem~\ref{thm:qary_multishot}(\ref{thm:strong_converse_multishot}) (strong converse)}\label{sec:proof strong multishot}
Consider a given sequence of $(n_i,l_i,2^{\Rni n_i^{l_i(q-1)}},\lambda_{1,i}, \lambda_{2,i})$ ID codes for  \bupcq  with  $n_i \rightarrow \infty$, $\Rni \rightarrow \infty$, and any sequence $l_i$.
Recall that $N_i$ is the number of type classes in $\cA_q^{n_i}$.
For each $i$,  using Lemma~\ref{lemma: permutation to noiseless_multishot}, we construct an $(N_i^{l_i}, 2^{\Rni n_i^{l_i(q-1)}},\lambda_{1,i}, \lambda_{2,i})$ ID code $\{(Q_j^{(i)},P_j^{(i)})|j=1,\ldots,2^{\Rni n_i^{l_i(q-1)}}\}$ for $\nlli$  with stochastic decoders. Rest of the proof of Theorem~\ref{thm:qary_multishot}(\ref{thm:strong_converse_multishot}) follows using the same steps as in the proof of Theorem~\ref{thm:oneshot}(\ref{thm:strong_converse}), and by setting 
\begin{align}
    \Rni'=\frac{\Rni}{4l_i e^{\beta(q-1)^2} (q-1)\log n_i } \label{eq:\Rni'_multishot}
\end{align}
in \eqref{eq:\Rni'}.

\section{Proof of Theorem~\ref{thm:feedback_deterministic}} \label{sec:feedback}
To prove the converse, consider an arbitrary $n,l\geq 1$, such that there exists a $(n,\numblocksfixed,M,\lambda_{1},\lambda_{2})$ IDF code $\{(\mbQ_{k,1}, \cdots, \mbQ_{k,\numblocksfixed},\cD_k)|k=1,\ldots,M\}$. If for any $k$, $\cD_k=\emptyset$, then clearly, $\lambda_{k \not\rightarrow k}= 1$ and this proves the result. Also $\cD_k$s are distinct, since otherwise, if there exists $j \neq k \in [M]$ such that $\cD_k=\cD_{j}$, then $\lambda_{k \not\rightarrow k}= 1-\lambda_{k \rightarrow j}$, which implies that $\lambda_{1}+\lambda_{2}\geq \lambda_{k \not\rightarrow k}+\lambda_{k \rightarrow j}=1 $.
Hence $M$ cannot be larger than the number of distinct non-empty subsets of $\cAq^{n\numblocksfixed}$, i.e.,
\begin{align*}
    & M < 2^{q^{n\numblocksfixed}}.
\end{align*}
 This completes the proof of the converse.

We now prove the achievability under deterministic IDF codes. We present a feedback coding scheme with deterministic encoder and decoders. Consider some arbitrary positive $R<1$, $l> \frac{2}{1-R}$ and an arbitrary sequence  $n_i \rightarrow \infty$.
The achievability scheme consists of two phases: the first is the common randomness (CR) generation phase of  $(\numblocksfixed-1)$ blocks (i.e. $n_i(\numblocksfixed-1)$ transmissions), followed by the message phase of $1$ block (i.e. $n_i$ transmissions). In the first phase, the transmitter establishes a common randomness of size $q^{n_i (\numblocksfixed-1)}$ (i.e. $n_i(\numblocksfixed-1) \log q$ bits) and in the second phase, the transmitter uses this common randomness and the message to select a sequence for transmission. The detailed scheme is described below.

Consider a fixed $i$, denote $n=n_i, M=M_i$; and describe the construction of a deterministic $(l,M)$ IDF code for \bupcnq. Recall that $\cTn_P$ denotes the type class of type $P$. Let us define 
\begin{align}
    P^* \coloneqq \underset{P \in \cTqn}{\text{argmax}} ~~|\cTn_P|,\notag 
\end{align}
as the type with the maximum number of vectors in its type class. 
Let $\vx^*$ be an $n$ length vector with type $P^*$. 
When $\vx^*$ is transmitted over \bupcnq\!, the channel output vector is uniform over $\cTn_{P^*}$. Note that by \eqref{eq:Nbounds},  we have, for $n\geq q-1$,
\begin{align} \label{eq:qary typeclass bound}
    \frac{ q^n }{(2n)^{q-1}} \le |\cTn_{P^*}| \le q^n. 
\end{align} 
Suppose $\vx_1,\vx_2,\cdots,\vx_N$ are a set of representatives 
of $N$ distinct types  $T_1,T_2,\cdots, T_N$ in $\cAq^n$ respectively.
We construct a family of functions 
\begin{align}
    \Phi_i : \cAq^{n(\numblocksfixed-1)} \rightarrow \{\vx_1,\vx_2,\cdots,\vx_N\}
\end{align}
for $i \in [M]$ by selecting the output of the mappings $\Phi_i(\cdot)$s independently and  uniformly at random i.e.,
$$
\mathbb{P}(\Phi_i(\uvy)= \vx_{j})=\frac{1}{N},
$$
for all $j \in [N]$, $\uvy \in \cAq^{n(\numblocksfixed-1)}, i \in [M]$. 

Let $\vx^{(j)}$ and $\vy^{(j)}$ denote respectively the $n$-length transmitted and received vectors in block $j$.
Our $(l,M)$ IDF encoder outputs the fixed vector $\vx^{(j)}=\vx^*$ in each of the first $\numblocksfixed -1$ blocks, irrespective of the message $i$. On receiving the corresponding output vectors $\vy^{(1)},\cdots, \vy^{(\numblocksfixed-1)}$ through feedback, the encoder outputs $\Phi_i(\vy^{(1)},\cdots ,\vy^{(\numblocksfixed-1)})$ in the last (i.e. the $l$-th) block if the message is $i$.
In other words, the IDF code is $\{(\vf_{i,1}, \ldots, \vf_{i,\numblocksfixed},\cD_i)|i=1, \ldots, M\}$, where for $j\in [\numblocksfixed-1]$ $\vf_{i,j}:\cAq^{2n(j-1)} \rightarrow \cAq^n$ are given by
\begin{align*}
    & \vf_{i,j}(\vx^{(1)}, \ldots, \vx^{(j-1)}, \vy^{(1)},\cdots, \vy^{(j-1)} )=\vx^*, \forall \vx^{(1)}, \ldots, \vx^{(j-1)}, \vy^{(1)},\cdots ,\vy^{(j-1)}\in \cAq^n; \forall j\in[\numblocksfixed -1]  \\
    &\vf_{i,\numblocksfixed} (\vx^{(1)}, \ldots, \vx^{(\numblocksfixed-1)}, \vy^{(1)},\cdots, \vy^{(\numblocksfixed-1)} )=\Phi_i(\vy^{(1)},\cdots, \vy^{(\numblocksfixed-1)}), \forall \vx^{(1)}, \ldots, \vx^{(\numblocksfixed-1)}, \vy^{(1)},\cdots, \vy^{(\numblocksfixed-1)}\in \cAq^n, 
\end{align*}
and 
\begin{align}
    \cD_i &\coloneqq \left\{(\vy^{(1)},\cdots, \vy^{(\numblocksfixed)})| \vy^{(1)},\cdots, \vy^{(\numblocksfixed-1)}\in \cT_{P^*}^{(n)}, \vy^{(\numblocksfixed)} \in \cT^{(n)}_{\Phi_i(\vy^{(1)},\cdots ,\vy^{(\numblocksfixed-1)})}\right\}.\notag
\end{align}
Here recall that $\cT_\vx^{(n)}$ denotes the typeclass containing $\vx$.

{\it{Analysis:}}
For any $0<R<1$, $l> \frac{2}{1-R}$, and $ n_i \rightarrow \infty $, we now prove that the sequence of $(l,2^{q^{Rln_i}})$ IDF code $\{(\vf_{j,1}, \ldots, \vf_{j,\numblocksfixed},\cD_j)|j=1, \ldots, 2^{q^{Rln_i}}\}$, obtained from the above construction, gives a sequences of $(n_i,l,2^{q^{Rln_i}},0,\lambda_{2,i})$ IDF codes over \bupcq such that $\lambda_{2,i} \rightarrow 0$ as $i \rightarrow \infty$.
Fix an $i$ and denote $n=n_i,$ and $\uvy=(\vy^{(1)},\cdots, \vy^{(\numblocksfixed-1)})$. The missed detection probability for message $j$ is given by 
\begin{align}
    \lambda_{j \not\rightarrow j} 
    &= \sum_{(\vy^{(1)},\ldots, \vy^{(\numblocksfixed)}) \in \cD_j^c} \left(\prod_{s=1}^{s=\numblocksfixed-1} \bfpcq (\vy^{(s)}|\vx^{*})\right)\bfpcq (\vy^{(\numblocksfixed)}|\Phi_j(\uvy)) =0 \notag 
\end{align}
for all  $j \in [2^{q^{Rnl}}]$. Hence 
\begin{align}
    \lambda_{1}=\max_{j \in [2^{q^{Rnl}}]}\lambda_{j \not\rightarrow j}=0.
\end{align}

Given any pair $j\neq k \in [2^{q^{Rnl}}]$ and $\uvy \in \cAq^{n(\numblocksfixed-1)}$, we define an indicator random variable $\ajk(\uvy)$ as
\begin{align*}
    \ajk(\uvy) = \begin{cases}
        1 \text{, if } \Phi_j(\uvy)=\Phi_k(\uvy),\\
        0 \text{, otherwise.}
    \end{cases}
\end{align*}
For any two messages $j \neq k$,  
we can write
\begin{align*}
    \lambda_{j \rightarrow k} 
    &= \sum_{(\vy^{(1)},\ldots, \vy^{(\numblocksfixed)}) \in \cD_k} 
    \left(\prod_{s=1}^{s=\numblocksfixed-1} \bfpcq (\vy^{(s)}|\vx^{*})\right)\bfpcq ( \vy^{(\numblocksfixed)}|\Phi_j(\uvy)) \\
    &= \sum_{\uvy \in (\cTn_{P^*})^{\numblocksfixed-1}} \frac{1}{|\cT_{P^*}^{(n)}|^{\numblocksfixed-1}}\sum_{\vy^{(l)} \in \cT^{(n)}_{\Phi_k(\uvy)} }   \bfpcq (\vy^{(\numblocksfixed)}|\Phi_j(\uvy)) \\
    &=  \frac{1}{|\cT_{P^*}^{(n)}|^{\numblocksfixed-1}}\sum_{\uvy \in (\cT_{P^*}^{(n)})^{\numblocksfixed-1}}  \ajk (\uvy).
\end{align*}
Since $\lambda_2=\max_{j \neq k}\lambda_{j \rightarrow k}$, we have
\begin{align}
\mathbb{P}\left( \lambda_2 \le \frac{2}{N}  \right)
    &= \mathbb{P}\left( \bigcap_{j \neq k}  \Bigg\{\frac{1}{|\cT_{P^*}^{(n)}|^{\numblocksfixed-1}}\sum_{\uvy \in (\cT_{P^*}^{(n)})^{\numblocksfixed-1}}  \ajk (\uvy) \le \frac{2}{N}  \Bigg\} \right) \notag \\
    &=1-\mathbb{P}\left( \bigcup_{j \neq k}\Bigg\{  \frac{1}{|\cT_{P^*}^{(n)}|^{\numblocksfixed-1}}\sum_{\uvy \in (\cT_{P^*}^{(n)})^{\numblocksfixed-1}}  \ajk (\uvy) > \frac{2}{N}  \Bigg\} \right) \notag \\
    &\ge 1-2^{2q^{Rnl}}  \times \mathbb{P}\left( \frac{1}{|\cT_{P^*}^{(n)}|^{\numblocksfixed-1}}\sum_{\uvy \in (\cT_{P^*}^{(n)})^{\numblocksfixed-1}}  \a12 (\uvy) > \frac{2}{N}  \right).\label{eq:Mbound}
\end{align}
Note that $\a12(\uvy);~ \uvy \in (\cT_{P^*}^{(n)})^{\numblocksfixed-1}$ are i.i.d. with mean $1/N$. Using Chernoff-Hoeffding Bound~\cite[Theorem 1]{hoeffding1994probability}, we have
\begin{align*}
    \mathbb{P}\left( \frac{1}{|\cT_{P^*}^{(n)}|^{\numblocksfixed-1}}\sum_{\uvy \in (\cT_{P^*}^{(n)})^{\numblocksfixed-1}}  \a12 (\uvy)\ge \frac{2}{N} \right)
    &\le 2^{ -|\cT_{P^*}^{(n)}|^{\numblocksfixed-1}\cD_{KL}(2/N||1/N)}
    \stackrel{}{\le} 2^{ -2|\cT_{P^*}^{(n)}|^{\numblocksfixed-1} (1/N)^2 \log e}.
\end{align*}
The last inequality uses Pinsker's inequality.
for large enough $n$, using \eqref{eq:Nbounds1} and \eqref{eq:qary typeclass bound} we can write 
\begin{align}
    \mathbb{P}\left( \frac{1}{|\cT_{P^*}^{(n)}|^{\numblocksfixed-1}}\sum_{\uvy \in (\cT_{P^*}^{(n)})^{\numblocksfixed-1}}  \a12 (\uvy)\ge \frac{2}{N} \right)
    &\stackrel{}{\le}2^{-
    \frac{2q^{n(l-1)}\log e}{\left((2n)^{(q-1)(l-1)} \times (2n)^{2(q-1)}\right)}
    }
    \stackrel{}{=}2^{-
    \frac{2q^{n(l-1)}\log e}{\left(2n\right)^{(q-1)(l+1)} }
    }\label{eq:LD bound}.
\end{align}
Therefore 
\begin{align}
\mathbb{P}\left( \lambda_2 \le \frac{2}{N}  \right)
    & \ge
    1-2^{2q^{Rnl}} \times  2^{-
    \frac{2q^{n(l-1)}\log e}{\left(2n\right)^{(q-1)(l+1)} }
    } \notag & \\
    &=1-2^{-2q^{nl}\left(
    \frac{\log e}{q^{\left( n+(q-1)(l+1)\log_q(2n)\right)}}-\frac{1}{q^{(1-R)nl}}
    \right)} \notag & \\
    &>0,  &\left(\text{ for large enough $n$, and $l > \frac{2}{1-R}$}\right) \notag 
\end{align} 
because for large enough $n$ and  $l> \frac{2}{1-R}$,  we have 
$n+(q-1)(l+1)\log_q(2n) < nl(1-R)$, if $R<1$.
    Hence there exists an   $(n,l,2^{q^{Rnl}},0,2/N)$  IDF code over \bupcnq for large enough $n$ and  $l> \frac{2}{1-R}$. 
    This completes the proof of  Theorem~\ref{thm:feedback_deterministic}(\ref{thm:ach_f}) because $ 2/N \rightarrow 0$, as $n \rightarrow \infty$.

\section{Conclusion}\label{sec:conclusion}

We studied identification over uniform permutation channels. Without feedback, our achievability result showed that message size growing as $2^{\epsilon_n n^{q-1}}$, where $\epsilon_n\rightarrow 0$, is identifiable. We showed a soft converse, proving that for any $R>0$, there do not exist ID codes with message size $2^{Rn^{q-1}}$  and power-law decay of the probability of error. We proved a strong converse that for any sequence of ID codes with $2^{R_nn^{q-1}}$ messages, where $R_n \rightarrow \infty$, the sum error probability converges to at least $1$. 
On the other hand, in the presence of noiseless and block-wise feedback, the identifiable number of messages may grow doubly exponentially as $2^{q^{Rnl}}$ for any $R<1$, when coding is done over $l$ blocks.
Therefore, in the binary case, feedback helps to increase the number of identifiable messages via permutation channels from sub-exponential to doubly exponential.

\section{Acknowledgements}
The work was supported in part by the Bharti Centre for Communication at IIT Bombay.

\bibliographystyle{unsrt}
\bibliography{bibfile_ID}

\appendices

\section{Proof of Lemma~\ref{lemma:delta_n lower bound}} \label{sec:pf_by_johnson}

We present an alternate proof of Lemma~\ref{lemma:delta_n lower bound} using Johnson's bound.

\noindent
\setcounter{lemma}{5}
\begin{lemma}[restated] 
    Let $\alpha \in (0,1)$  and $n$ be fixed. 
    Consider a collection of subsets of $[N]$,  $\cU=\{\cU_1,\cdots,\cU_M\}$ with set sizes $N\epsilon$ and maximum pairwise intersection $N\delta$.
    If $M > \frac{N}{\alpha}$ 
    then
    $\delta >(1-\alpha)\epsilon^2$.
\end{lemma}
\begin{IEEEproof}
    Let $A_2(N,d,w)$ be the maximum number of codewords of a binary code of length $N$, minimum Hamming distance $d$, and constant codeword weight $w$. Johnson bound~\cite{huffman2010fundamentals} gives that
if $2w^2-2Nw+Nd>0$, then
    \begin{align}
        A_2(N,d,w) \le \frac{Nd}{2w^2-2Nw+Nd}.
    \end{align}
We will now prove that for a given collection of subsets $\cU$ of $[N]$, if $\delta \leq (1-\alpha)\epsilon^2$, then $M\leq \frac{N}{\alpha}$. 
The characteristic vectors of the subsets in $\cU$ form an $N$ length constant weight binary code with
\begin{align*}
    d&=2N(\epsilon-\delta),\\
    w&=N\epsilon.
\end{align*}
Since $\delta \leq (1-\alpha)\epsilon^2$, this implies $\epsilon^2>\delta$, and hence
\begin{align*}
    &2w^2 - 2Nw +Nd \\
= & 2N^2\epsilon^2 - 2N^2\epsilon + 2N^2 (\epsilon - \delta)\\
= & 2N^2 (\epsilon^2 - \delta) > 0
\end{align*}
Hence by Johnson's bound, we have
\begin{align*}
    M  & \le \frac{Nd}{2w^2 - 2Nw +Nd}\\
    & \le \frac{2N^2(\epsilon-\delta)}{2N^2 (\epsilon^2 - \delta)}\\
    &=\frac{\epsilon-\delta}{\epsilon^2-\delta}\\
    &\le \frac{\epsilon}{\epsilon^2-\delta}\\
    &\stackrel{(a)}{\le } \frac{\epsilon}{\epsilon^2-(1-\alpha)\epsilon^2}\\
    &=\frac{1}{\alpha\epsilon}\\
    &\stackrel{(b)}{\le}\frac{N}{\alpha}.
\end{align*}
Here $(a)$ follows because $\delta \le (1-\alpha)\epsilon^2$, $(b)$ follows because $\epsilon \ge \frac{1}{N}$. 
This proves the lemma.
\end{IEEEproof}

\section{Proof Of Lemma~\ref{lemma: permutation to noiseless_multishot}} \label{app:proof permutation to noiseless_multishot}

We will  construct an ID code $\{(Q'_i, P_i) | i=1,\ldots, M\}$ with stochastic decoders for $\nll$ in two steps: 
First, we construct an ID code $\{(Q'_i,P_i)|i=1, \ldots, M\}$ with a stochastic decoder for the channel 
$\left(NL_{[N]} \right)^l$ with same error probabilities. Next, we argue that since there is a bijection $\phi:[N^l] \rightarrow [N]^l$  and both 
$\left(NL_{[N]}\right)^l$  and 
$\nll$ are noiseless channels, the same ID code $\{(Q'_i,P_i)|i=1, \ldots, M\}$ can also be used for the channel 
$\nll$, which results in the same error probabilities.

Let $T_1,T_2,\cdots, T_N$ be the distinct types for vectors in $\cAqn$.
For every type $T \in \cTqn$, recall that $\cTn_{T} \subset \cAqn$ denotes the \emph{type class} containing all $n$-length vectors with type $T$.
For every $i\in [M]$, $j_s \in [N]$ for all $s \in [l]$, we define
    \begin{align*}
        Q'_{i}(j_1, \ldots,j_l) &\coloneqq
        Q_i(\cTn_{T_{j_1}}\times \cTn_{T_{j_2}}\times \ldots \times \cTn_{T_{j_l}})=
        \sum_{\vx^{(1)} \in \cTn_{T_{j_1}}; \ldots; \vx^{(l)} \in \cTn_{T_{j_l}} } Q_i(\vx^{(1)}, \ldots, \vx^{(l)})\\
        P_i(1|j_1, \ldots, j_l)&\coloneqq \frac{|\cD_i \cap \cTn_{T_{j_1}}\times \cTn_{T_{j_2}}\times \ldots \times \cTn_{T_{j_l}}| }{|\cTn_{T_{j_1}}\times \cTn_{T_{j_2}}\times \ldots \times \cTn_{T_{j_l}}|}.
    \end{align*}
    The probability of errors for the new code are given by
    \begin{align*}
        \tilde{\lambda}_{i \rightarrow j} &\coloneqq \sum_{(j_1, \ldots,j_l) \in [N]^l}Q'_i(j_1, \ldots,j_l)P_j(1|j_1, \ldots,j_l)\\
        &=\sum_{(j_1, \ldots,j_l) \in [N]^l}\sum_{\vx^{(1)} \in \cTn_{T_{j_1}}; \ldots; \vx^{(l)} \in \cTn_{T_{j_l}} } Q_i(\vx^{(1)}, \ldots, \vx^{(l)}) \frac{|\cD_i \cap \cTn_{T_{j_1}}\times \cTn_{T_{j_2}}\times \ldots \times \cTn_{T_{j_l}}| }{|\cTn_{T_{j_1}}\times \cTn_{T_{j_2}}\times \ldots \times \cTn_{T_{j_l}}|}\\
        &=\sum_{(\vx^{(1)}, \ldots, \vx^{(l)}) \in \cAq^{nl}}Q_i(\vx^{(1)}, \ldots, \vx^{(l)}) \frac{|\cD_i \cap \cTn_{\vx^{(1)}}\times \cTn_{{\vx^{(2)}}}\times \ldots \times \cTn_{{\vx^{(l)}}}| }{|\cTn_{{\vx^{(1)}}}\times \cTn_{{\vx^{(2)}}}\times \ldots \times \cTn_{{\vx^{(l)}}}|}=\lambda_{i \rightarrow j}.
    \end{align*}
    Following a similar argument, we can show that
    \begin{align}
         \lambda_{i \not\rightarrow i}
         =\tilde{\lambda}_{i \not\rightarrow i}.
    \end{align}
    This proves the lemma.

\end{document}